\documentclass[prb,twocolumn,letterpaper]{revtex4}
\usepackage{dcolumn}
\usepackage[colorlinks=true,citecolor=blue,urlcolor=blue]{hyperref}
\usepackage{longtable}
\usepackage{graphicx}
\def\be{\begin{equation}}
\def\ee{\end{equation}}
\begin{document}

\bibliographystyle{apsrev}
\title{\boldmath Interpretation of the in-plane infrared response
  of the high-${T_{\rm c}}$ cuprate superconductors 
  involving spin fluctuations revisited \unboldmath} 
\author{Petr C\'{a}sek$^{1}$}
\author{Christian Bernhard$^{2}$}
\author{Josef Huml{\'\i}{\v c}ek$^{1}$}
\author{Dominik Munzar$^{1}$} 
\email{munzar@physics.muni.cz}
\affiliation{$^{1}$Institute of Condensed Matter
  Physics, Faculty of Science, Masaryk University,
  Kotl{\'a}{\v r}sk{\'a} 2, 61137 Brno, Czech Republic\\
$^{2}$Max-Planck-Institut f{\"u}r Festk{\"o}rperforschung,
Heisenbergstrasse 1, D-70569 Stuttgart, Germany}

\begin{abstract}
The in-plane infrared response 
of the high-$T_{\rm c}$ cuprate superconductors 
was studied using the spin-fermion model, 
where charged quasiparticles of the copper-oxygen planes 
are coupled to spin fluctuations.  
First, we analyzed structures 
of the superconducting-state conductivity 
reflecting the coupling of the quasiparticles to the resonance mode  
observed by neutron scattering. 
The conductivity $\sigma$ computed with the input spin susceptibility 
in the simple form of the mode  
exhibits two prominent features:
an onset of the real part of $\sigma$  
starting around the frequency $\omega_{0}$ of the mode 
and a maximum of a related function $W(\omega)$, 
roughly proportional to the second derivative of the scattering rate 
$[1/\tau](\omega)$,
centered approximately at $\omega=\omega_{0}+\Delta_{0}/\hbar$, 
where $\Delta_{0}$ is the maximum value of the superconducting gap. 
The two structures are well known from earlier studies.  
Their physical meaning, however, 
has not been sufficiently elucidated thus far.  
Our analysis involving quasiparticle spectral functions 
provides a clear interpretation.
Second, we explored the role played by the spin-fluctuation continuum,  
whose  spectral weight 
is known to be much larger than the one of the mode. 
We have shown that the experimental spectra of $1/\tau$
can be approximately reproduced 
by augmenting the resonant-mode component of the spin susceptibility 
by a suitable continuum component with a considerably higher spectral weight
and with a characteristic width of several hundreds meV. 
The computed spectra of $1/\tau$ display 
a new structure in the mid-infrared 
which is related to the finite width of the occupied part 
of the conduction band. 
Third, we investigated the temperature dependence (TD) of $\sigma$ 
assuming that the normal state spin susceptibility consists 
of an overdamped low energy mode and the continuum component.  
The differences between the experimental normal-state spectra 
and those of the superconducting state, 
including some interesting effects at higher frequencies, 
are reasonably well reproduced. 
Motivated by recent experimental (ellipsometric) works  
by Molegraaf and coworkers 
[H.~J.~A.~Molegraaf {\it et al.}, Science {\bf 295}, 2239 (2002).]
and Boris and coworkers 
[A.~V.~Boris {\it et al.}, Science {\bf 304}, 708 (2004).], 
we further studied the TDs
of the effective kinetic energy ${\rm K.E.}$ 
and of the intraband spectral weight $I_{\rm O}$.   
Calculations for the trivial case 
of noninteracting quasiparticles in the normal state 
and a BCS-like superconducting state 
reveal a strong sensisitivity 
of the TD of $I_{\rm O}$ to details 
of the dispersion relation. 
The TDs of ${\rm K.E.}$ and $I_{\rm O}$ in the interacting case, 
for the set of the values of the input parameters used throughout this work,  
are similar to those of the trivial case.
The physics beyond the changes occuring 
when going from the normal to the superconducting state, 
however, is shown to be more complex, 
involving, besides the formation of the gap, 
also a feedback effect of the spin fluctuations 
on the quasiparticles 
and a significant shift of the chemical potential. 
\end{abstract}

\maketitle

\section{\label{sec:int} Introduction}

Models of the electronic structure 
of the high-$T_{\rm c}$ cuprate superconductors (HTCS), 
where charged quasiparticles of the copper-oxygen planes 
are coupled to spin fluctuations (SF)
(for review see Refs.~\onlinecite{norman03,abanov03})
possess two appealing features.
(a) They provide a straightforward explanation 
of the symmetry of the order parameter and 
(b) the same value of the coupling constant leads both 
to a fairly good agreement between theory and experiment 
for the normal state 
and to values of $T_{c}$ of $50-100{\rm\,K}$. 
 
On the other hand, these models suffer two important deficiencies. 
(a) They cannot be rigorously derived 
starting from a well established microscopic hamiltonian 
such as the hamiltonian of the 2D one-band Hubbard model. 
They can only be motivated by some perturbation expansions 
of the latter.  
Within the so called conserving fluctuation exchange (FLEX) approximation 
\cite{bickers89,monthoux94B,pao94,dahm95,scalapino95,lichtenstein96,manske01}, 
e.g., 
an important term in the expression 
for the quasiparticle selfenergy 
$\Sigma({\bf k},{\rm i}\omega_{n})$ reads 
\begin{equation}
{1\over \beta N}{3U^{2}\over 2}\sum_{{\mathbf q},{\rm i}\Omega_{m}}
\chi_{\rm RPA}({\mathbf q},{\rm i}\Omega_{m})
G({\mathbf k}-{\mathbf q},{\rm i}\omega_{n}-{\rm i}\Omega_{m})\,,
\label{eq:Selfen1}
\end{equation}
where $i\omega_{n}$ ($i\Omega_{m}$) are the fermion (boson) Matsubara frequencies, 
$\beta=1/(k_{B}T)$, $N$ is the number of momentum points, and 
$U$ is the Hubbard parameter. 
Further, $G$ is the dressed quasiparticle propagator and 
$\chi_{RPA}$ the random-phase-approximation(RPA)-based expression 
for the spin susceptibility.  
The expression for the selfenergy used within the SF-based models 
has precisely the same structure:
\begin{equation}
{1\over \beta N}{3g^{2}\over 4}\sum_{{\mathbf q},{\rm i}\Omega_{m}}
\chi_{\rm SF}({\mathbf q},{\rm i}\Omega_{m})
G({\mathbf k}-{\mathbf q},{\rm i}\omega_{n}-{\rm i}\Omega_{m})\,,
\label{eq:Selfen2}
\end{equation}
where $g$ and $\chi_{\rm SF}$ are the model coupling constant 
and spin susceptibility, respectively. 
This can be considered as a kind of formal justification 
of the SF-based approaches. 
A typical value of $g$  
of $0.5{\rm\,eV}$, however,  
is an order of magnitude lower than that of $U$ 
resulting from first-principles calculations
of  $5-10{\rm\,eV}$ \cite{mcmahan88,hybertsen89}.
(b) The second problem of the SF-based models concerns  
the choice of the function $\chi_{\rm SF}$. 
In the spirit of the models, 
$\chi_{\rm SF}$ should match the true spin susceptibility $\chi$ 
accessible to experimental investigations.  
Unfortunately, the available experimental data are not yet sufficient 
to determine $\chi$ in the relevant spectral range. 
This applies especially to the normal state (NS)
of optimum doped and overdoped materials, 
where the spin fluctuations are relatively weak.  
The form of $\chi_{\rm SF}$ has thus to be guessed, 
at least to some extent. 
In the classical series of papers, Monthoux {\it et al.} 
\cite{monthoux92,monthoux93,monthoux94A}
have adopted the Millis-Monien-Pines formula\cite{millis90}: 
\begin{equation}
\chi_{\rm MMP}({\bf q},\omega)=
{\chi_{Q}\over {1+({\bf q} -{\bf Q})^{2}\,\xi^{2}
- {\rm i}\omega/\omega_{\rm sf}}}\,, 
\label{eq:MMP}
\end{equation}
where ${\bf Q}=(\pi/a,\pi/a)$ is the antiferromagnetic wave vector, 
$a$ is the in-plane lattice parameter, 
$\xi$ the antiferromagnetic coherence length, and 
$\omega_{sf}$ is the parameter that specifies 
the frequency of the relaxation mode 
(at a fixed~${\bf q}$, $\chi''_{\rm MMP}({\bf q},\omega)$ 
exhibits a broad maximum centered at 
$\omega=\omega_{\rm sf}[1+({\bf q} -{\bf Q})^{2}\,\xi^{2}]$, 
i.e., a relaxation mode). 

The situation is simpler for the superconducting state (SCS), 
where a sharp resonance in $\chi$ (``magnetic mode") 
centered at ${\bf q}={\bf Q}$, 
${\bf Q}=(\pi/a,\pi/a)$, 
and $\omega=\omega_{0}$, 
$\omega_{0}\approx 40{\rm\,meV}$, 
has been detected in neutron scattering experiments 
(see Refs.~\onlinecite{fong96},\onlinecite{fong99},
\onlinecite{fong00},\onlinecite{he02} 
and references therein).
For some compounds the shape of the resonance is well known 
and it is thus possible to investigate 
the coupling between the mode and the charge carriers
without any {\it ad hoc} assumptions regarding the form of $\chi_{\rm SF}$. 
Let us emphasize that there must be a coupling 
between the well defined mode and 
the well defined Bogoliubov quasiparticles and 
it should be considered 
in any theory of superconductivity in the HTCS, in particular, 
when describing single-particle excitations (observed by photoemission)
and two-particle excitations (observed, among others, by optics).
Such considerations may help to clarify 
the role played by the SF 
in the mechanism of superconductivity. 
The single-particle case has been reviewed in detail 
in Refs.~\onlinecite{norman03} and \onlinecite{eschrig03}.
An important observation is 
that the peak-dip-hump structure observed  
in superconducting-state angle-resolved-photoemission spectra 
(for review see Ref.~\onlinecite{damascelli03})
at and around $(\pi/a,0)$ can be understood and described 
in terms of the above mentioned coupling.
This result, together with several other findings,  
suggests that the SF are an important player in the mechanism.  
In this work we focus on the influence of the coupling 
on the in-plane infrared response of the HTCS in the superconducting state. 

The SF-based approach has been pioneered in this context 
by Schachinger, Carbotte, and Marsiglio \cite{schachinger97}    
and by Quinlan, Hirschfeld, and Scalapino \cite{quinlan96}. 
The former group used 
the boson spectral density proportional to  
$\chi''_{\rm MMP}({\bf q}={\bf Q},\omega)$, 
in the work of the latter $\chi_{\rm RPA}$ has been used. 
The real part of the in-plane infrared conductivity, $\sigma_{1}(\omega)$, 
resulting from these computations exhibits the following trends: 
(i) Below $T_{c}$ and for frequencies lower than $\sim 4\Delta_{0}$, 
where $\Delta_{0}$ is the maximum value of the superconducting gap,  
it decreases with decreasing temperature. 
(ii) It does not possess a true energy gap, 
even at very low temperatures and in the absence of any impurity scattering.   
Instead, at low temperatures, it increases gradually with increasing $\omega$. 
(iii) The increase becomes steeper above the characteristic 
spin-fluctuation frequency 
of $\omega_{\rm sf}=30{\rm\,meV}$\cite{schachinger97} 
or $\Delta_{0}$ - $2\Delta_{0}$\cite{quinlan96}. 
(iv) A very broad maximum appears 
in the low-temperature spectra around $\sim 4\Delta_{0}$.  
These findings are roughly consistent with experimental data  
\cite{puchkov96,bernhard02,tu02,vdmarel03,homes04,boris04}, 
except for two features. 
First, the data contain additional sharp structures at low frequencies. 
These are probably related to charge inhomogeneities 
and will not be discussed here.  
Second, the increase of $\sigma_{1}(\omega)$ in the data 
appears to be less gradual than in the theoretical spectra. 
The data rather seem to exhibit 
an onset starting around $300-400{\rm\,cm^{-1}}$. 
This may be related to the presence of a sharp feature 
in the spin-fluctuation spectra below $T_{c}$, 
i.e., to the magnetic mode, 
which has not been considered 
in Refs.~\onlinecite{schachinger97} and \onlinecite{quinlan96}. 

The mode has been taken into account  
by two of the authors and M.~Cardona (MBC)\cite{munzar99}, 
who employed   
a form of $\chi_{\rm SF}$ 
reflecting the results of the neutron scattering experiments:  
\begin{equation}
\chi_{\rm RM}({\bf q},\omega)=
 {1\over 
 {1+({\bf q} -{\bf Q})^{2}\,\xi^{2}}}\,
 {F\over {\omega_{0}^{2}-\omega^{2}-{\rm i}\Gamma\omega}}\,, 
\label{eq:MBC}
\end{equation}
where $\Gamma$ is the broadening of the resonance mode 
and~$F$ expresses its ``oscillator strength". 
This study provided a tentative interpretation of the onset feature: 
the final states consist, in the first approximation,  
of two Bogoliubov quasiparticles and the magnetic mode;
the minimum excitation energy is thus 
$\hbar\omega_{0}\approx 40{\rm\,meV}\approx 320{\rm\,cm^{-1}}$
and the broad maximum around $1000{\rm\,cm^{-1}}$ corresponds 
to excitations involving quasiparticles around the saddle point 
and the mode. 
Carbotte, Schachinger, and Basov (CSB)\cite{carbotte99}
also interpreted the data in terms of the magnetic mode 
but an inverse strategy has been used. 
Their approach is based on a relation 
between the conductivity 
and the electron-phonon spectral density $\alpha^{2}F(\omega)$
that applies to the normal state of a weakly coupled isotropic
electron-phonon system \cite{Schulga}:
\begin{equation}
  \alpha^{2}F(\omega)\approx W(\omega)=
{\varepsilon_{0}\omega_{\rm pl}^{2}\over 2\pi}
  {{\rm d}^{2}\over {\rm d}\omega^{2}}
  \left[\omega
{\rm Re}{1\over \sigma(\omega)}\right]\,,  
  \label{eq:wom1}
\end{equation}
where $\omega_{\rm pl}$ is the plasma frequency of the free charge carriers. 
CSB assumed (and to some extent verified by computing $\sigma(\omega)$ 
within Eliashberg formalism) that a slightly modified equation,
\begin{equation} 
I^{2}\chi_{ef}(\omega-\Delta_{max})\approx W(\omega)=
{\varepsilon_{0}\omega_{pl}^{2}\over 2\pi}
{{\rm d}^{2}\over {\rm d}\omega^{2}}
\left[\omega
{\rm Re}{1\over \sigma(\omega)}\right]\,,  
  \label{eq:wom2}
\end{equation}
can be used to obtain the 
``effective electron-spin-fluctuation spectral function" 
$I^{2}\chi_{ef}(\omega)$ of a $d$-wave superconductor \cite{schachinger97}.  
The main point of CSB is that $\chi_{ef}(\omega)$, 
as determined from the optical data,   
contains a sharp structure that is  
fairly similar to the one of $\chi''({\bf Q},\omega)$, i.e., 
to the neutron resonance. 
Based on this finding, CSB concluded that the infrared data  
reflect a coupling of the charge carriers to the magnetic mode. 
The coupling strength inferred from experiment 
is found to be sufficient to account for the high value of $T_{c}$. 
The relation between $\sigma_{1}(\omega)$ and the magnetic mode
has been further explored by Abanov, Chubukov, and Schmalian (ACS) 
\cite{abanov01A,abanov01B}. 
For example, they predicted a sharp onset 
of $\sigma_{1}(\omega)$ located 
at $\hbar\omega=2\Delta_{0}+\hbar\omega_{0}$ 
(see Fig.~2 of Ref.~\onlinecite{abanov01A}).  
The three approaches (MBC, CSB, ACS) differ in their emphasis 
and computational formalism.  
MBC showed that the superconducting-state infrared data can be understood 
in terms of the SF-based model with the spin susceptibility 
in the form of the neutron peak 
and focused on the onset in the conductivity spectra 
starting, according to the model, around 
$\omega_{0}$.
CSB found an approximate solution of the inverse problem, 
the spectral function $W(\omega)$,  
and attributed its main maximum 
to the magnetic mode shifted by $\Delta_{0}$ towards higher frequencies.
Finally, ACS concentrated on the contribution to the conductivity
of the so called hot spots, i.e., 
intercepts of the Fermi surface 
and the antiferromagnetic-Brillouin-zone (BZ) boundary,  
and they argued that this contribution is the dominant one. 
Regarding the formalism, all the three groups expressed 
the quasiparticle selfenergy using the SCS version of 
Eq.~\ref{eq:Selfen2}
and the conductivity using the textbook formulas \cite{schrieffer},  
where the vertex corrections are neglected.
The differences consist in treating the propagator $G$ 
and the susceptibility $\chi_{\rm SF}$. 
In the work of MCB, $G$ is replaced with the bare Nambu matrix
containing an estimated superconducting gap of $d_{x^{2}-y^{2}}$ symmetry, 
i.e., with the BCS expression. The approach is thus not self-consistent. 
On the other hand, it has two important advantages, 
not shared by the other two approaches,
besides its formal simplicity:  
(a) It is not restricted to models, 
where the SF are the only cause of superconductivity. 
(b) The full $q$-dependence of $\chi$ is taken into account. 
The same approach has been used 
in the studies of the single-particle properties 
by Eschrig and Norman and coworkers 
(see Ref.~\onlinecite{eschrig03} and references therein) 
and, in the context of electron-phonon interaction, by Sandvik, 
Scalapino, and Bickers \cite{sandvik04}.  
CSB obtain the propagator $G$ by solving Eliashberg equations 
with a separable interaction, i.e., in a selfconsistent way. 
ACS use an approximate form of these equations 
valid in a region around the hot spots. 
Not only the propagator but also the spin susceptibility 
is treated in a self-consistent way. 

The structures in the conductivity spectra 
can be relatively easily related to and understood in terms of 
those of the quasiparticle spectral functions 
(in the absence of vertex corrections).  
Such an analysis, however, has not yet been performed,
except for some observations regarding the contributions of the hot spots
\cite{munzar99,abanov01B} 
and the optical scattering rate \cite{schachinger03}.  
One of the aims of the present paper is to provide 
a detailed interpretation 
of the onset starting around $\omega_{0}$ 
discussed by MBC 
and of the main maximum of the function $W(\omega)$
emphasized by CSB
in terms of the quasiparticle spectral function. 
Among others, we explore the role played by the hot spots. 

The structure of the paper is the following. 
In Sec.~\ref{sec:teo} we summarize the basic equations of our approach 
and in Sec.~\ref{sec:cdt} we discuss the input parameters 
and present some computational details.  
Section \ref{sec:res} contains our results. 
We start with the quasiparticle spectral functions, 
that are compared with those computed by Eschrig and Norman 
\cite{eschrig03}
(Subsec.~\ref{ssec:akr}),  
and contributions of the individual $\mathbf{k}$-points 
to the real part of the conductivity 
(Subsec.~\ref{ssec:kpt}). 
Subsecs.~\ref{ssec:icd} and \ref{ssec:dmW} deal 
with the spectra of the total conductivity, 
in particular with its onset around $\omega_{0}$, 
and with the structures of the scattering rate and the function $W(\omega)$, 
respectively. 
In Subsec.~\ref{ssec:cont} we comment on the role of the SF continuum. 
Some aspects of the temperature dependence (TD) of the spectra, 
as the role of the TD of $\chi_{SF}$,  
and the TD of the integrated spectral weight, 
that has been recently related 
to changes of the effective in-plane kinetic energy
\cite{molegraaf02,boris04}, 
are addressed in Subsec.~\ref{ssec:TD}. 
A summary and our conclusions are presented in Sec.~\ref{sec:cnc}. 
Readers who are interested only in the main findings of the paper 
can skip Secs.~\ref{sec:teo} and \ref{sec:cdt}
and some more technical parts of Sec.~\ref{sec:res}. 

\section{\label{sec:teo} Theoretical framework}

Our starting point is the Hamiltonian of the spin-fermion model: 
\begin{equation}
H=H_0+H_{\rm int}\,,
\label{eq:Ham1}
\end{equation}
where 
\begin{equation}
H_0=\sum_{\mathbf{k}\alpha}\epsilon_\mathbf{k}
c^+_{\mathbf{k}\,\alpha}c_{\mathbf{k}\,\alpha}\,,  
\label{eq:Ham2}
\end{equation}
\begin{equation}
  H_{\rm int}=g \sum_{\mathbf{q}} {\bf s}(\mathbf{q} )\cdot
  {\bf S}(-\mathbf{q})\,. 
  \label{eq:Ham3}
\end{equation}
Here
$\epsilon_{\mathbf{k}}$ is the quasiparticle dispersion, 
$\alpha$ is the spin index, 
$g$ is the spin-fermion coupling constant, 
\begin{equation}
\mathbf{s}(\mathbf{q})=\frac{1}{\sqrt{N}}\frac{1}{2}
\sum_{\mathbf{k} \alpha\beta}
c_{\mathbf{k}+\mathbf{q} \alpha}^{+}
\mathbf{\tau}_{\alpha \beta}
c_{\mathbf{k} \beta}
\label{eq:elspin}
\end{equation}
is the Fourier component of the electronic spin, 
$\mathbf{\tau}=(\tau_{1},\tau_{2},\tau_{3})$ 
is the vector of the Pauli matrices, 
and ${\bf S}(\mathbf{q})$ the spin-fluctuation operator. 
The latter quantity is defined so 
that its retarded Green's function 
is equal to the negatively taken spin susceptibility, 
\begin{equation} 
\chi_{ij}(\mathbf{q},\omega)=\frac{1}{\hbar}\int\limits _{-\infty}^\infty
\chi_{ij}(\mathbf{q},t){\rm e}^{{\rm i}\omega t}{\rm d}t
\label{eq:chi1}
\end{equation}
with 
\begin{equation}
\chi_{ij}(\mathbf{q},t)=
{\rm i}\Theta(t)\left<[s_i(\mathbf{q},t),s_j(-\mathbf{q},0)]\right>\,.
\label{eq:chi2}
\end{equation}
It is assumed that $\chi_{ij}(\mathbf{q},\omega)$ is isotropic, 
i.e., $\chi_{ij}(\mathbf{q},\omega)= 
\delta_{ij}\chi(\mathbf{q},\omega)$.  

\subsection{\label{ssec:ak} Quasiparticle selfenergy and spectral function}

We follow here the approach of Ref.~\onlinecite{munzar99}  
and express the quasiparticle selfenergy $\Sigma({\mathbf k},iE_{n})$  
using second order perturbation theory ($H_{int}$ being the perturbation) 
and starting from a BCS state 
with an estimated superconducting gap of $d_{x^{2}-y^{2}}$ symmetry. 
The resulting formula for $\Sigma({\mathbf k},iE_{n})$ (2 by 2 matrix)  
represents a SCS version of Eq.~\ref{eq:Selfen2}:  
$$\Sigma(\mathbf{k},iE_n)=$$
\begin{equation}
=
{1\over \beta N}{3g^{2}\over 4}\sum_{{\mathbf q}}
\sum_{i\Omega_{m}}
\chi_{SF}({\mathbf q},i\Omega_{m})
G_{0}({\mathbf k}-{\mathbf q},iE_{n}-i\hbar\Omega_{m})
\label{eq:Selfen3}
\end{equation} 
with 
\begin{equation}
  G_{0}(\mathbf{k},iE_n)=
  \frac{{\rm i}E_n\tau_0+(\epsilon_{\mathbf{k}}-\mu)\tau_3+
    \Delta_{\mathbf{k}}\tau_1}{({\rm i}E_n)^2-
    (\epsilon_{\mathbf{k}}-\mu)^2-\Delta_{\mathbf{k}}^2}.
  \label{eq:GBCS}
\end{equation}
Here 
$G_{0}$  
is the bare Nambu Green's function 
${\rm i}E_{n}$ are the fermion Matsubara energies 
(${\rm i}E_{n}={\rm i}\hbar\omega_{n}$), 
$\mu$ is the chemical potential, 
and $\Delta_{\mathbf k}$ the superconducting gap. 
The full Nambu propagator $G$ 
is given by the Dyson equation, 
$G^{-1}=G_{0}^{-1}-\Sigma^{-1}$, and the matrix spectral function 
$A(\mathbf{k},E)=-2\,\mathrm{Im}\left[G_{ret}(\mathbf{k},E)\right]$ by 
\begin{equation}
A(\mathbf{k},E)=
-2\,\mathrm{Im} \left[G_0^{-1}(\mathbf{k},{\rm i}E_n)-
  \Sigma(\mathbf{k},{\rm i}E_n)\right]^{-1}
  _{{\rm i}E_n\rightarrow E+{\rm i}\delta}.
  \label{eq:Spectralfc}
\end{equation}

Equation \ref{eq:Selfen3} can be further rewritten into a form 
which is more suitable for practical computations: 
\begin{widetext}
\begin{equation}
\Sigma(\mathbf{k},{\rm i}E_n)=
{1\over N}{3g^{2}\over 4}\sum_{\mathbf q}
\int_0^\infty
{\rm d}\omega\hbar B(\mathbf{q},\omega)\,
  \left\{\frac{1}{2}(T_+ + T_-)\tau_0 +
             \frac{\epsilon_{\mathbf{k}+\mathbf{q}}}
             {2E_{\mathbf{k}+\mathbf{q}}}
             (T_{+}-T_{-})\tau_3 + 
             \frac{\Delta_{\mathbf{k}+\mathbf{q}}}{2E_{\mathbf{k}+\mathbf{q}}}
             (T_+ - T_-)\tau_1\right\}\,,
  \label{eq:sumT}
\end{equation}
\end{widetext}
where
\begin{equation}
  T_{\pm} = \frac{N_{\rm B}(\hbar\omega)+n_{\rm F}(\pm
    E_{\mathbf{k}+\mathbf{q}})}
    {iE_n+\hbar\omega\mp E_{\mathbf{k}+\mathbf{q}}} 
  + \frac{N_{\rm B}(\hbar\omega)+1-n_{\rm F}(\pm
    E_{\mathbf{k}+\mathbf{q}})}
   {iE_n-\hbar\omega\mp E_{\mathbf{k}+\mathbf{q}}}.
  \label{eq:tpm}
\end{equation}
In the preceding equations 
\begin{equation}
  B(\mathbf{q},\omega)=\frac{1}{\pi}
  \mathrm{Im}\{\chi_{\rm SF}(\mathbf{q},\omega)\} 
  \label{eq:Spectralfcsf}
\end{equation}
is the spectral function of the spin fluctuations. 
The quasiparticle excitation energies~$E_{\mathbf{k}}$ are defined 
as usual:
$E_{\mathbf{k}}=\sqrt{(\epsilon_{\mathbf{k}}-\mu)^2+\Delta_{\mathbf{k}}^2}$,
$N_{\rm B}$ and $n_{\rm F}$~are the Bose-Einstein and Fermi-Dirac distribution
functions, respectively.
For a single CuO$_{2}$ plane 
$(1/N)\sum_{\mathbf q}$ in Eqs.~\ref{eq:Selfen3},\ref{eq:sumT}
can be replaced with 
$\int_{\rm 2D\,BZ}a^{2}{\rm d}^{2}{\mathbf q}/ (2\pi)^{2}$.

\subsection{\label{ssec:or} Optical response}

The Kubo formula for the conductivity 
~$\sigma_{ij}(\mathbf{q},\omega)$ ($i$, $j$ $\in\{x,y,z\}$)
reads \cite{Mahan1}: 
\begin{equation}
\sigma_{ij}(\mathbf{q},\omega)=
\frac{{\rm i}}{\omega+{\rm i}\delta}
\left[R_{ij}(\mathbf{q},\omega)
+\frac{n_0e^2}{m}\delta_{ij}\right],
\label{eq:kubo}
\end{equation}
where
\begin{equation}
  R_{ij}(\mathbf{q},\omega)=-\frac{{\rm i}}{\hbar}\int\limits_{-\infty}^\infty
  {\rm d}t\theta(t){\rm e}^{{\rm i}\omega\,t}\left<\left[j_i(\mathbf{q},t),
      j_j(-\mathbf{q},0)\right]\right> 
\label{eq:korij} 
\end{equation}
is the retarded correlation function 
of the paramagnetic current density operator
and $n_{0}$ is the average electron density. 
The two terms in the bracket on the right hand side of Eq.~\ref{eq:kubo} 
[multiplied by the factor ${\rm i}/(\omega+{\rm i}\delta)$]
are the paramagnetic and the diamagnetic term, respectively. 
The contribution to $\sigma_{ij}(\mathbf{q},\omega)$ 
of charge carriers in a single band is given by a formula \cite{scalapino92}
which has the same structure as Eq.~\ref{eq:kubo}. 
The Fourier component of the corresponding current density operator 
is given by
\begin{equation}
{\mathbf j}({\mathbf q})=
{e\over \sqrt{V}}
\sum_{\mathbf{k}\alpha}
{\mathbf v}
\left({\mathbf k}+{{\mathbf q}\over 2}\right)
c^{+}_{\mathbf{k}\,\alpha}c_{\mathbf{k}+\mathbf{q}\,\alpha}\,,
\label{eq:joneb}
\end{equation}
where 
${\mathbf v}({\mathbf k})=(1/\hbar)\partial \epsilon/\partial {\mathbf k}$,
and the diamagnetic term is replaced with
\begin{equation}
-\frac{\rm i}{\omega+{\rm i}\delta}
{e^{2}\over \hbar^{2}}\langle K_{ij}\rangle \,,\, 
K_{ij}=-{1\over V}
\sum_{\mathbf{k}\alpha}
{\partial^{2} \epsilon\over \partial k_{i}\partial k_{j}}
c^+_{\mathbf{k}\,\alpha}c_{\mathbf{k}\,\alpha}\,.  
\label{eq:diamt}
\end{equation}

For easy reference, we summarize here some basic properties 
of the  optical conductivity 
$\sigma_{ij}(\omega)=\sigma_{ij}(\mathbf{q}=0,\omega)$. 
It consists of a singular part $\sigma_{ij}^{s}(\omega)$, 
related to the condensate, 
and a regular part $\sigma_{ij}^{r}(\omega)$.
They are given by
\begin{equation} 
\sigma^{s}(\omega)=
{{\rm i}\varepsilon_{0}\omega_{\rm pl,sc}^{2}\over \omega+i\delta}
\label{eq:sigs}
\end{equation}
and 
\begin{equation}
\sigma^{r}(\omega)={{\rm i}\over\omega}
\left[R(\omega)-{\rm Re}\{R(0)\}\right]\,. 
\label{eq:sigr}
\end{equation}
The tensor indices are omitted for simplicity, 
$\omega_{\rm pl,sc}$ is the plasma frequency of the superfluid, 
\begin{equation}
\omega_{\rm pl,sc}^{2}=
-{1\over \varepsilon_{0}}
\left[
{e^{2}\over \hbar^{2}}\langle K\rangle -
{\rm Re}\{R(0)\}
\right]
\label{eq:omplsc1}
\end{equation}
and $R(\omega)=R(\mathbf{q}=0,\omega)$.
The imaginary part $\sigma_{2}(\omega)$ of $\sigma(\omega)$ 
can be expressed in terms of the real part, $\sigma_{1}(\omega)$, 
using the following Kramers-Kronig relation:   
\begin{equation}
\sigma_{2}(\omega)=
-{1\over \omega}{e^{2}\over \hbar^{2}}\langle K\rangle+
{2\over \omega\pi}
{\rm P}\int_{0^{+}}^{\infty}{\rm d}{\omega'}
{\omega'^{2}\sigma_{1}(\omega')-\omega^{2}\sigma_{1}(\omega)
\over \omega^{2}-\omega'^{2}}\,. 
\label{eq:KKtr}
\end{equation} 
The symbol $0^{+}$ in the lower limit means 
that the singular component of $\sigma_{1}$,  
$\sigma^{s}_{1}(\omega)=
\pi\varepsilon_{0}\omega_{\rm pl,sc}^{2}\delta(\omega)$,  
is excluded from the integration. 
The function $\sigma_{1}(\omega)$ (including its singular component) 
satisfies the following useful sum rule:  
\begin{equation}
I_{O}=\int_{0}^{\infty}{\rm d}\omega \sigma_{1}(\omega)=
{\pi\over 2}\varepsilon_{0}\omega_{\rm pl}^{2}=
-{\pi\over 2}{e^{2}\over \hbar^{2}}\langle K\rangle\,,    
\label{eq:sumr}
\end{equation}
which allows us to express the plasma frequency of the superfluid as 
\begin{equation}
\omega_{\rm pl,sc}^{2}=\omega_{\rm pl}^{2}-
{2\over \pi\varepsilon_{0}}
\int_{0^{+}}^{\infty}{\rm d}\omega \sigma_{1}(\omega)\,. 
\label{eq:omplsc2}
\end{equation} 

The correlation function $R_{ij}(\omega)$, 
which is the key quantity of the theory,
can be obtained
by the analytical continuation of the corresponding Matsubara Green's
function $P_{ij}(\mathbf{q},i\Omega_{m})$. 
Using the Wick's theorem and neglecting vertex corrections we obtain
\cite{Schrieffer2}:
$$P_{ij}(\mathbf{q},i\Omega_m)=$$
\begin{equation}
= -2e^{2}
  {1\over V}\sum_{\mathbf k}
  v_i\left(\mathbf{k} +\frac{\mathbf{q}}{2}\right)v_j\left(\mathbf{k}
    +\frac{\mathbf{q}}{2}\right) L(\mathbf{k},\mathbf{q},{\rm i}\Omega_m),
  \label{eq:pij}
\end{equation}
where
$L(\mathbf{k},\mathbf{q},{\rm i}\Omega_m)=$
\begin{equation}
=-\frac{1}{2\beta}\sum\limits_{n=-\infty}^\infty
{\rm Tr}[G(\mathbf{k} +\mathbf{q},
{\rm i}E_n+{\rm i}\hbar\Omega_m)G(\mathbf{k},{\rm i}E_n)]\,.
\label{eq:lgg}
\end{equation}
The regular component of $\sigma_{ij\,1}(\omega)$ 
can be expressed in terms of the matrix spectral function 
of Eq.~\ref{eq:Spectralfc} 
as follows \cite{Mahan2}:
\begin{widetext}
\begin{equation}
\sigma_{ij\,1}(\omega)=
\frac{e^2}{2\omega}
{1\over V}\sum_{\mathbf k}
v_{i}\left(\mathbf{k}\right)v_{j}\left(\mathbf{k}\right)
\int\frac{{\rm d}E}{2\pi}{\rm Tr}
\left[A(\mathbf{k},E)A(\mathbf{k},E+\hbar\omega)\right]\times
\left[n_{\rm F}(E)-n_{\rm F}(E+\hbar\omega)\right]\,.  
\label{eq:cond}
\end{equation}
\end{widetext}
For a superconductor containing  
$N_{p}$ equivalent and weakly coupled CuO$_{2}$ planes 
within a unit cell, 
$(1/V)\sum_{\mathbf k}$ in Eqs.~\ref{eq:diamt}, \ref{eq:pij}, \ref{eq:cond}
can be replaced with 
$(N_{p}/d)\int_{\rm 2D\,BZ}{\rm d}^{2}{\mathbf k}/(2\pi)^{2}$, 
where $d$ is the lattice parameter along the $c$-axis.  

\section{\label{sec:cdt} Input parameters and computational details}

\subsection{\label{ssec:dsp} Dispersion relation}

\begin{figure}[htbp]
  \centering
  \includegraphics[width=0.45\textwidth]{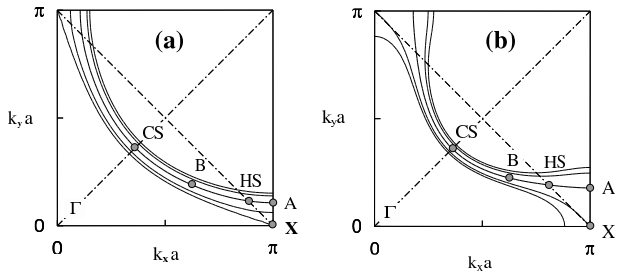}  
  \caption{Fermi surface (thick line) for the 
    dispersion given by Eq.~\ref{eq:dsp} (a) and for the dispersion
    of Ref.~\onlinecite{eschrig03} (b). Supplementary contours (thin lines)
    correspond to excitation energies of $\pm 0.033$~eV and $\pm
    0.050$~eV. A notation of some special points is introduced.  The
    point~CS (``cold spot'') is the crossing point of the Fermi
    surface and the BZ~diagonal (dashed-dotted line starting in the
    $\Gamma$~point). The point~HS (``hot spot'') is the crossing point
    of the Fermi surface and the antiferromagnetic BZ~boundary
    (dashed-dotted line starting in the X~point). The point~B is
    located between the~CS and the~HS and its (dimensionless)
    coordinates are $(1.96,0.60)$ and (1.96,0.78) in (a) and (b), respectively. }
  \label{fig:dsp}
\end{figure}

In this work we use 
a tight-binding expansion of the in-plane dispersion relation 
including the second nearest neighbor hopping terms: 
\begin{equation}
  \varepsilon(\mathbf{k})=
  -2t[\cos(k_{x}a)+\cos(k_{y}a)]-4t'\cos(k_{x}a)\cos(k_{y}a)\,.  
  \label{eq:dsp}
\end{equation}
The values of the parameters are the same as in Ref.~\onlinecite{munzar99}
and they are summarized in Table \ref{tab:ft}.  
\begin{table*}[Htbp]
  \centering
  \setlength{\tabcolsep}{6pt}
  \begin{tabular}{cccccccccc}
    \hline \hline \\[2 mm]
    $a$(\AA) & $d$(\AA) & $t$ (eV) & $t'$ (eV) & $\mu$ (eV) &
    $\Delta_{0}$ (eV) &
    $g$ (eV) & $\hbar\omega_0$ (eV) & $\Gamma$ (eV) & $\xi$ (\AA) \\[2 mm]
    \hline\\[2 mm] 
   3.828 & 11.650 & 0.250 & -0.100 & -0.350 & 0.030 & 0.350 & 
   0.040 & 0.010/0 & 9.0 (=2.35$a$) \\[2 mm]
    \hline \hline 
  \end{tabular}
  \caption{Values of the parameters used in the computations.
    The values of $a$ and $d$ correspond to YBa$_{2}$Cu$_{3}$O$_{7-\delta}$. 
    We consider two CuO$_{2}$ planes within a unit cell, 
    as in YBa$_{2}$Cu$_{3}$O$_{7-\delta}$, i.e., $N_{p}=2$.}
  \label{tab:ft}
\end{table*}
The corresponding Fermi surface is shown in part (a) of Fig.~\ref{fig:dsp}. 
For comparison, the Fermi surface, 
corresponding to the six-parameter fit of the dispersion relation 
of optimally doped Bi$_{2}$Sr$_{2}$CaCu$_{2}$O$_{8}$ (Bi-2212)
used in Ref.~\onlinecite{eschrig03}, is shown in part (b). 
The figure is further used to introduce 
a notation of some important ${\mathbf k}$-points.
Within the SF-based scenarios, 
the quasiparticles around the CS (HS) are weakly (strongly) renormalized 
by the spin-fermion coupling. 
The CS coincides with the node of the superconducting gap 
whereas the HS is located 
in the region of high values of $|\Delta|$. 

Our approach will be applied here to the bilayer compounds, 
i.e., materials, that contain two CuO$_{2}$ planes within a unit cell. 
The conduction band of these compounds can be expected to be splitted 
into the bonding branch and the antibonding branch \cite{andersen95}.
For a long time, it has not been clear, 
whether such splitting indeed occurs 
or is supressed by strong correlations in the CuO$_{2}$ planes. 
Two well defined bands were recently observed 
in photoemission experiments on strongly overdoped Bi-2212 
and some signatures of the splitting have been reported 
also for optimally doped and underdoped Bi-2212 
\cite{feng01,chuang01,kordyuk02} 
(for a review see Ref.~\onlinecite{damascelli03}).
It has been proposed that some of the effects associated earlier 
with the magnetic mode, 
in particular the peak-dip-hump structure of the photoemission spectra, 
can be understood solely in terms of the two bands. 
Subsequently it has been shown, however, 
that a ``true", i.e., linked to selfenergy effects, 
peak-dip-hump structure 
develops in the spectra below $T_{c}$ \cite{eschrig02,borisenko03}. 
In the present study, 
the bonding-antibonding splitting is not taken into account,   
i.e., two identical conduction bands per unit cell 
are assumed, 
corresponding to two equivalent and independent planes. 
It is possible to go beyond this approximation \cite{lichtenstein96,eschrig02,li05}
but we believe that it would not seriously modify our conclusions 
regarding the in-plane infrared spectra.

\subsection{\label{ssec:gap} Superconducting gap}

If not specified in the text, we use a common ansatz for the
superconducting gap of $d_{x^2-y^2}$-symmetry:
\begin{equation}
\Delta(\mathbf{k})=\frac{\Delta_0}{2}[\cos (k_xa)-\cos (k_ya)]
\label{eq:gap} 
\end{equation}
with $\Delta_{0}=30{\rm\,meV}$, 
a value close to the experimental result 
for optimally doped Bi-2212 \cite{damascelli03}.
The coupling to the spin-fluctuations slightly enhances  
the value of the gap in the region around the HS.  
For the value of $g$ of $350{\rm\,meV}$ used in our computations, 
the renormalized value of $|\Delta|$ exactly at the hot spot 
is about $35{\rm\,meV}$. 
This fact, however, 
has no significant impact on the quantities discussed. 

\subsection{\label{ssec:ssc} Spin susceptibility and the coupling constant}
In most of the computations 
we use the spin susceptibility 
of the resonance-mode form of Eq.~\ref{eq:MBC} 
with the values of the input parameters given in Table \ref{tab:ft}. 
The FWHM of the resonance peak $\Gamma$ is small
as compared to both~$\omega_{0}$ and~$\Delta_{0}$. 
It can be therefore neglected, in the
first approximation, 
an approach applied by Eschrig and Norman \cite{eschrig03}. 
We have performed the calculations for  
both $\Gamma=10{\rm\,meV}$ and $\Gamma=0$, i.e.,  
\begin{equation}
  \label{eq:sss}
  B (\mathbf{q},\omega) =
  \frac{F\pi}{1+\left(\mathbf{q}-\mathbf{Q}\right)^2 \xi^2}\;
  \delta(\omega-\omega_0)\,, 
\end{equation}
in order to identify the role of the finite linewidth. 
In the following we shall refer to the calculations with 
$\Gamma=10{\rm\,meV}$ and $\Gamma=0$ as to those 
with the broad and the sharp magnetic resonance, respectively. 
The value of the ``oscillator-strength" parameter $F$  
is determined by using the same normalization condition 
as in Ref.~\onlinecite{munzar99}:
\begin{equation}
  I_{\rm M}=\hbar\int\frac{a^2{\rm d}^2q}{(2\pi)^2}\int_0^\infty {\rm d}\omega 
                        \left(2\,
    N_{\rm B}(\hbar\omega) + 1\right)\,B(\mathbf{q},\omega) = \frac{1}{4}\,,
  \label{eq:normalizationchi}
\end{equation}
where $B(\mathbf{q},\omega)$~is the spectral function 
of the spin fluctuations 
defined by Eq.~\ref{eq:Spectralfcsf}.
Equation \ref{eq:normalizationchi} represents 
an approximate ``total momentum sum rule" 
(see Ref.~\onlinecite{fong00} for a discussion).  

Note that when using  
$\chi_{\rm SF}$ of the resonance-mode form  
together with Eq.~\ref{eq:normalizationchi} 
we make two approximations. 
(a) Within the model of two identical conduction bands 
per bilayer unit cell
the effective spin susceptibility 
is equal to the average of the so called odd ($\chi_{\rm o}$) 
and even ($\chi_{\rm e}$) component 
(for a definition see Ref.~\onlinecite{fong00}). 
Here we take into account only the resonant part of $\chi_{\rm o}$,  
i.e., we do not consider $\chi_{\rm e}$ 
and the continuum part of $\chi_{\rm o}$. 
This may be a reasonable starting point 
for an analysis of the low-energy phenomena. 
First, the magnitude of $\chi_{\rm o}$ 
is typically by a factor of 2-3 larger than that of $\chi_{\rm e}$ 
\cite{fong00,pailhes03}, 
and second, 
the continuum part 
cannot be expected to cause any sharp structure 
in the quasiparticle selfenergy 
(see Ref.~\onlinecite{eschrig03} for a detailed discussion). 
(b) In Eq.~\ref{eq:normalizationchi} 
the resonance is assumed to collect one half of the total 
magnetic spectral weight, $I_{\rm M}=1/4$.  
Experimentally, the value of $I_{\rm M}$ in Eq.~\ref{eq:normalizationchi} 
corresponding to the mode 
is only about $0.01 \times (1/4)$ 
for optimally doped YBa$_{2}$Cu$_{3}$O$_{7-\delta}$ (Y-123) \cite{fong00}
and only ca $0.04 \times (1/4)$ 
for optimally doped Bi-2212 \cite{fong99}. 
Since $F\sim I_{\rm M}$ enters Eq.~\ref{eq:Selfen2} 
in the product $g^{2}F$, however, 
we continue to use Eq.~\ref{eq:normalizationchi} for simplicity
keeping in mind 
that the actual value of the coupling constant  
required  
is an order of magnitude higher than the one of Table \ref{tab:ft}. 

Let us make here a short comment on a recent discussion 
regarding the small experimental values of $I_{\rm M}$.
Kee, Kivelson, and Aeppli \cite{kee02} claimed that they are too small 
for the mode to play any significant role.  
Abanov {\it et al.} \cite{abanov02} in their response to the criticism argued 
that even though $I_{\rm M}$ is small, 
the mode can lead to large selfenergy effects because 
it is rather narrow in ${\mathbf q}$ space. 
There is another argument justifying the ``mode models". 
As mentioned above $F\sim I_{\rm M}$ appears   
in Eq.~\ref{eq:Selfen2} 
only in the product $g^{2}F$. 
The susceptibility $\chi_{\rm RM}$ normalized to 1/4, 
as in our calculations,      
together with $g=0.35{\rm\,eV}$, 
will be shown to lead to results 
that are to some extent in agreement with experimental data.    
The same results could also be obtained using  
``the true $\chi_{\rm RM}$", 
i.e., $\chi_{\rm RM}$ normalized 
to the experimental value of $0.002$ ($0.009$), 
with $g\approx 4{\rm\,eV}$ ($g\approx 2{\rm\,eV}$). 
The important point is that 
the latter values of $g$ are not unrealistic given  
Eq.~\ref{eq:Selfen1} and the high values of $U$ 
expected to be relevant for the HTCS.
It is thus well possible that, 
inspite of the small values of $I_{\rm M}$, 
the selfenergy effects 
due to the mode are large. 
The fact, however, that its spectral weight is small 
as compared to that of the SF continuum  
implies that only some low energy features can be attributed to the mode.  
The intermediate energy-scale spectra must be 
determined rather by the continuum. 

\subsection{\label{ssec:compd} Computational details}

The results presented in this work have been obtained using two different codes.
In the first one, the finite linewidth of the resonance is taken into account,  
the second one is adapted to the case of the sharp resonance. 
The calculations proceed essentially in three steps. 
 
(i) In the first step, 
the self-energies and spectral functions 
of selected $\mathbf{k}$-points 
are computed by using Eqs.~\ref{eq:sumT} and \ref{eq:Spectralfc}. 
First we concentrate on the frequency integration of Eq.~\ref{eq:sumT}.  
Within the sharp resonance approach,   
it is trivial due to the $\delta$ function in Eq.~\ref{eq:sss}. 
The structures of the resulting function, however, are sharp
and the sampling of the subsequent BZ integration must be very dense. 
Within the broad resonance calculation,  
the integral of Eq.~\ref{eq:sumT} is divided into two parts:   
the part containing the Bose factors in the numerators of Eq.~\ref{eq:tpm}
and the one containing the remaining terms. 
The latter part can be expressed analytically but
the former one has to be calculated numerically. 
For this purpose, 
we have used the grid containing 500 $\omega$-points 
ranging from $\omega=0$ to $\omega=10k_{\rm B}T/\hbar$. 
Due to the separable form of $\chi_{RM}$,  
the same frequency integrals occur for any $\mathbf{k}\in 1.\,{\rm BZ}$. 
This has allowed us to compute their values  
once and for all at the beginning. 

Next, we focus on the BZ integration.  
The dispersion relation is nontrivial and 
the structure in the susceptibility relatively sharp.
In addition, a discussion of the singularities requires 
the spectral functions to be continued very close to the real axis: 
a typical value of $\delta$ in Eq.~\ref{eq:Spectralfc} is only $0.05{\rm\,meV}$. 
For all these reasons a very fine grid
(typically $300\times 300$ for the broad resonance 
and $1000\times 1000$ for the sharp one) 
is needed for the BZ~integration.
Without further simplifications the calculations of the infrared conductivity 
would be fairly time consuming: 
they would require elaborate integrations  
for all $\mathbf{k}$-points of the corresponding grid. 
This problem can be partially overcome by using an approximation 
described below. 
The BZ integral of Eq.~\ref{eq:sumT} can be schematically written as 
\begin{equation}
\int_{\rm 2D\,BZ}{\rm d}\mathbf{q}\,\zeta(\mathbf{q}) 
D(\mathbf{k}+\mathbf{q},iE_{n})\,,  
\label{eq:cd1}
\end{equation}
where $\zeta$ is the $\mathbf{q}$ dependent part of $B$ and 
$D$ is a function of $\mathbf{k}+\mathbf{q}=\mathbf{k'}$ and $iE_{n}$. 
Using the $\mathbf{k}$-space periodicity of both $\zeta$ and $D$ 
we obtain 
\begin{equation} 
\int_{\rm 2D\,BZ}=
\int_{\rm 2D\,BZ}{\rm d}\mathbf{k'}\,\zeta(\mathbf{k'}-\mathbf{k}) 
D(\mathbf{k'},iE_{n})\,.  
\label{eq:cd2}
\end{equation}
The idea underlying our approximation is that 
$\zeta$ is a smooth function as compared to $D$. 
This allows us to use the approximation  
\begin{equation}
\int_{\rm 2D\,BZ}\approx \sum_{ij}
\zeta(\mathbf{k}_{ij}-\mathbf{k})D_{ij}(iE_{n})\,,
\label{eq:cd3}
\end{equation} 
where $i=1,2,\,...\,N_{1}$, $j=1,2,\,...\,N_{1}$, 
$N_{1}$ specifies the number of $\mathbf{k}$-points  
in the auxiliary grid (typically $N_{1}=30$),  
$\mathbf{k}_{ij}=(k_{xi},k_{yi})$,  
$k_{xi}=(2i-1)\pi/(aN_{1})$, 
$k_{yj}=(2j-1)\pi/(aN_{1})$, and 
\begin{equation}
D_{ij}(iE_{n})=
\int_{2(i-1)\pi/(aN_{1})}^{2i\pi/(aN_{1})}
\int_{2(j-1)\pi/(aN_{1})}^{2j\pi/(aN_{1})}{\rm d}\mathbf{k}
D(\mathbf{k},iE_{n})\,.
\label{eq:cd4}
\end{equation}
The quantities $D_{ij}(iE_{n})$ 
have been computed once and for all at the beginning, 
using grids of typically $10\times 10$ ($30\times 30$) $\mathbf{k}$-points 
for the broad (sharp) resonance.  

(ii) In the second step, the contributions of selected $\mathbf{k}$-points
to the real part of the (isotropic) in-plane infrared conductivity 
are calculated. 
We define the contribution of a $\mathbf{k}$-point as 
(cf.~Eq.~\ref{eq:cond})
\begin{widetext}
\begin{equation}
\Delta \sigma_{1}(\mathbf{k},\omega)=
\frac{e^2}{2\omega}
{N_{p}\over da^{2}}\,
[v_{x}^{2}\left(\mathbf{k}\right)+v_{y}^{2}\left(\mathbf{k}\right)]
\int\frac{{\rm d}E}{2\pi}{\rm Tr}
\left[A(\mathbf{k},E)A(\mathbf{k},E+\hbar\omega)\right]\, 
\times \left[n_{\rm F}(E)-n_{\rm F}(E+\hbar\omega)\right]\,.  
\label{eq:ccond}
\end{equation}
\end{widetext}
The computation of the convolution 
of the spectral functions in Eq.~\ref{eq:ccond}
is again very demanding 
because the sharp quasiparticle peaks must be handled correctly. 
Typically a grid of at least 5000~energy-points 
for a spectrum ranging up to $300{\rm\,meV}$ is required. 
In the calculations of Subsecs.~\ref{ssec:cont} and \ref{ssec:TD}, 
where the sharp spectral structures are not at the centre of interest,
the spectral functions have been continued  
only to a distance of typically $1{\rm\,meV}$ from the real axis and  
energy grids of typically 6000 points for spectra ranging up to $3{\rm\,eV}$ 
have been used. 

(iii) The calculation of $\sigma_{1}(\omega)$ 
is completed by the BZ integration of $\Delta\sigma_{1}$:
\begin{equation}
\sigma_{1}(\omega)=4a^{2}\int_{\rm 2D\,IBZ}{\rm d}\mathbf{k}\,
\Delta \sigma_{1}(\mathbf{k},\omega)\,, 
\end{equation}
where IBZ stands for the irreducible part of the BZ. 
The integration is relatively simple 
because the functions $\Delta \sigma_{1}$ are already fairly smooth.
The grid of $200 \times 200$ $\mathbf{k}$-points in the full BZ,
i.e., ca 5000 $\mathbf{k}$-points in its irreducible part, 
proved to be sufficient.  

If not specified in the text, the computations are performed 
for $T=20{\rm\,K}$. 

\section{\label{sec:res} Results and discussion}

The infrared spectra will be analyzed  
in terms of the quasiparticle spectral functions. 
For this reason, we first present and discuss, 
as a necessary prerequisite,  
our calculated spectral functions 
of some representative $\mathbf{k}$-points (Subsec.~\ref{ssec:akr})
and their contributions to the real part of the conductivity 
(Subsec.~\ref{ssec:kpt}).  
The main new results of the paper 
will then be presented in Subsects.~\ref{ssec:icd}-\ref{ssec:TD}.  

\subsection{\label{ssec:akr} Quasiparticle spectral functions}

\begin{figure}[htbp]
  \centering
  \includegraphics[width=0.45\textwidth]{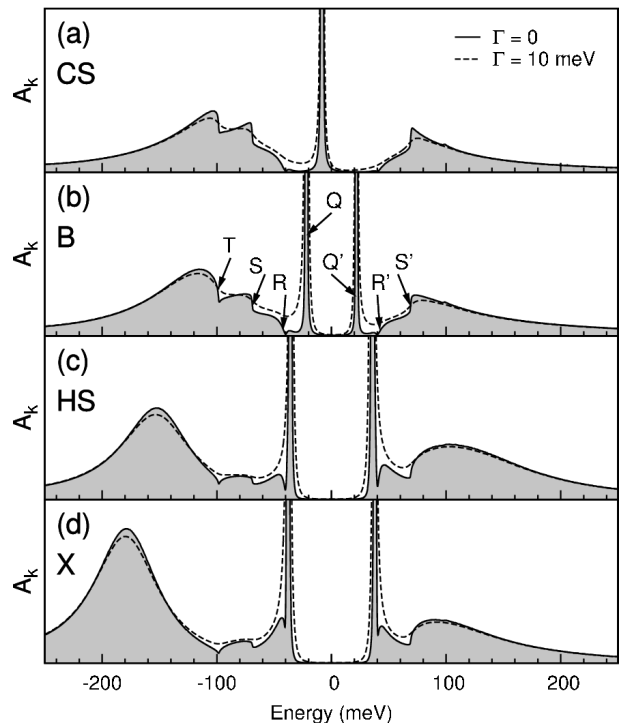}    
  \caption{Spectral functions of the $\mathbf{k}$-points
    defined in Fig.~\ref{fig:dsp}.  
    The notation of spectral structures is introduced in panel (b). }
  \label{fig:sp1}
\end{figure}
Figure~\ref{fig:sp1} shows the spectral functions 
(more precisely,  the first diagonal elements 
of the matrix (\ref{eq:Spectralfc}), 
i.e., the ``true'' spectral functions)
of three points located at the Fermi surface 
(the points CS, B, and HS) and of the X-point. 
The spectra exhibit quasiparticle peak(s) 
(one for the~CS and two for the other points,   
where the magnitude of the superconducting gap is finite), 
that would appear already for the (noninteracting) BCS ground-state, 
and incoherent broad satellites at higher energies,
that are due to the interaction.  
The structures are more pronounced in the spectra for $\Gamma=0$.  
The latter are thus more suitable 
for a discussion of the singularities. 

The energies of the quasiparticle peaks 
labelled Q and Q' 
(see Fig.~\ref{fig:sp1})  
are $-8{\rm\,meV}$
(CS, the nonzero value is due to a slight renormalization of the Fermi surface
caused by the interaction),  
$\pm 22{\rm\,meV}$ (B), $\pm 35{\rm\,meV}$ (HS), and $\pm 37{\rm\,meV}$ (X). 
The magnitude of the values for the HS is larger than $\Delta_{0}$ which  
reflects an enhancement of superconductivity 
due to the coupling to the mode. 

Next we discuss the incoherent parts of the spectral functions. 
For all of the $\mathbf{k}$-points,  
three onset features can be resolved at negative energies,  
denoted by R, S, and T (see Fig.~\ref{fig:sp1}), 
and two at positive energies (R', S').  
Interestingly, their positions are the same for all the $\mathbf{k}$-points: 
the feature R(R') is located at 
$-(+)\hbar\omega_{0}$ [$\mp 40{\rm\,meV}$], 
S(S') at 
$-(+)(\hbar\omega_{0}+\Delta_{A})
\approx -(+)(\hbar\omega_{0}+\Delta_{0})$
[$\approx \mp 70{\rm\,meV}$] 
and 
T at 
$-|\hbar\omega_{0}+E_{X}|$, where
$E_{X}=\sqrt{\epsilon_{X}^2+\Delta_{X}^{2}}$ 
[$\approx -100{\rm\,meV}$]. 
The physical meaning of the structures S,S', and T 
has been discussed in detail by Eschrig and Norman \cite{eschrig03};  
their analysis also yields the formulas for the energies of the features. 
The structures R and R', 
which are well resolved only for $\Gamma=0$,   
correspond to final states involving a nodal quasiparticle 
and the magnetic excitation.  
The energy of the broad maximum at negative energies, below the feature $T$, 
is $\mathbf{k}$-dependent. Its values are     
$ -100{\rm\,meV}$ (CS), 
$ -115{\rm\,meV}$ (B),  
$ -150{\rm\,meV}$ (HS), 
and $ -180$~meV (X). 
The decrease of the energy when going from the CS to the X-point 
is accompanied by a systematic increase 
of the spectral weight. 

\begin{figure}[htbp]
  \centering \includegraphics[width=0.45\textwidth]{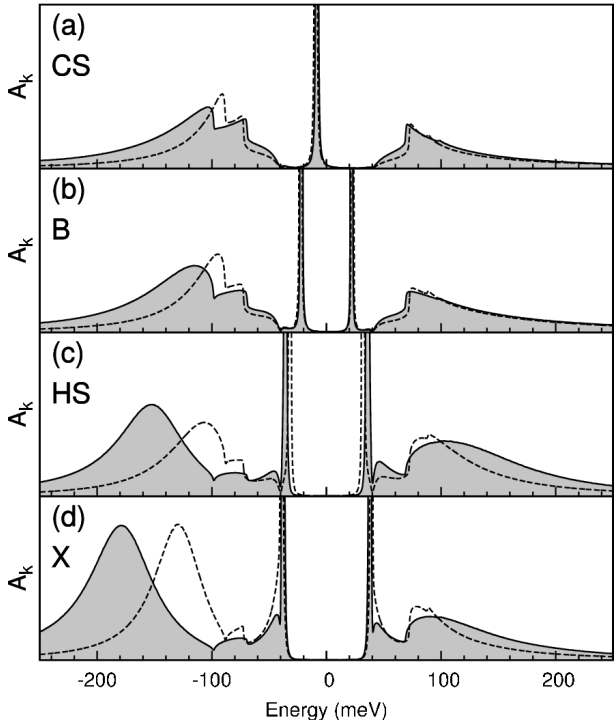}
  \caption{Spectral functions 
    from Fig.~\ref{fig:sp1} (solid lines) compared with those 
    obtained using the values of the input parameters 
    of Ref.~\onlinecite{eschrig03} (dashed lines).} 
  \label{fig:sp2}
\end{figure}

In order to illustrate the role of the input parameters, 
we compare in Fig.~\ref{fig:sp2} the results presented above 
with those obtained 
using the parametrizations of the dispersion relation 
and of the magnetic susceptibility  
and the values of $\Delta_{0}$ ($35{\rm\,meV}$), 
$g=0.65$, and $T$ ($40{\rm\,K}$) 
of Ref.~\onlinecite{eschrig03}.     
Note first that the reference spectra are in good agreement with 
those of Ref.~\onlinecite{eschrig03}. 
Note further, that for small absolute values of energy, 
the two sets of spectra are very similar. 
On the other hand, the positions and shapes 
of the lowest incoherent peak are fairly different. 
This is due to the fact 
that the value of the effective coupling constant 
used by Eschrig and Norman is considerably smaller than the ours \cite{eschrigg}. 
The difference in the energy of the onset of the latter peak 
(ca $ -100{\rm\,meV}$ in this work, 
ca $\ -90{\rm\,meV}$ in Ref.~\onlinecite{eschrig03}) 
is related to the difference in the location of the X-point 
with respect to the Fermi surface (see Fig.~\ref{fig:dsp}).

\subsection{\label{ssec:kpt} Contributions of the selected $\mathbf{k}$-points 
to $\sigma_{1}(\omega)$}

\begin{figure}[htbp]
  \centering \includegraphics[width=0.45\textwidth]{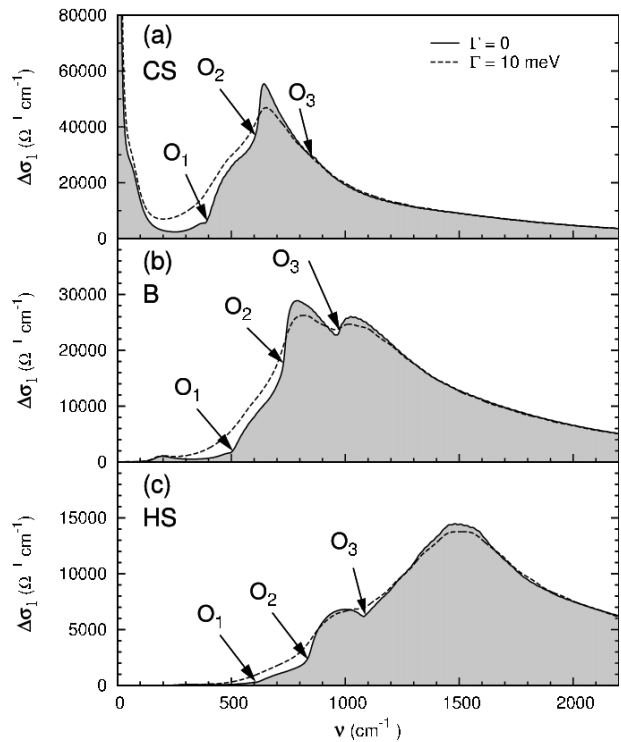}
  \caption{Contributions of the selected $\mathbf{k}$-points 
    to the real part of the optical conductivity.  
    The onset features are marked by the arrows.  
}
  \label{fig:cc1}
\end{figure}

The contributions of the $\mathbf{k}$-points labelled as CS, B, and HS 
to $\sigma_{1}(\omega)$
are shown in Fig.~\ref{fig:cc1}. 
The contribution of the X-point is identically zero because
$\mathbf{v}({\rm X})=0$.
In the following, the structures of the functions  
will be discussed in terms of those 
of the corresponding spectral functions. 
Such a discussion is straightforward only for the normal state, 
where the expression on the r.~h.~s.~of Eq.~\ref{eq:cond}
reduces to a convolution of the (true) spectral functions. 
For the superconducting state, 
the off-diagonal components 
of the spectral-function matrix (\ref{eq:Spectralfc})
play a crucial role. 
For example, 
the transitions ${\rm Q}\rightarrow{\rm Q'}$ do not contribute  
because of a precise cancellation 
of the term on the r.~h.~s.~of Eq.~\ref{eq:cond} 
containing the diagonal components 
and that containing the off-diagonal components. 
Other transitions between occupied and unoccupied states, 
however, do show up in the conductivity spectra, 
only their spectral weights are influenced 
by the presence of the off-diagonal components. 

Three singularities  
(discontinuities of the first derivative),   
can be observed in the spectra.  
In Fig.~\ref{fig:cc1} they are marked by the arrows. 
The onset feature ${\rm O}_{1}$   
corresponds to the transitions 
${\rm Q}\rightarrow{\rm R'}$ and ${\rm R}\rightarrow {\rm Q'}$
(for the CS, ${\rm Q}\equiv {\rm Q'}$). 
It appears at the energy of ${\bar E}_{\mathbf{k}}+\hbar\omega_{0}$, 
where ${\bar E}_{\mathbf{k}}$ is the energy of the quasiparticle peak.   
The values of the corresponding wavenumbers are   
$380{\rm\,cm}^{-1}$, $500{\rm\,cm}^{-1}$, and  
$600{\rm\,cm}^{-1}$ for the points CS, B, and HS, respectively. 
The second onset feature ${\rm O}_{2}$ 
is related to the transitions 
${\rm Q}\rightarrow{\rm S'}$ and ${\rm S}\rightarrow{\rm Q'}$. 
It is shifted with respect to $O_{1}$ approximately by $\Delta_{0}$,  
the values of the corresponding wavenumbers are   
$620{\rm\,cm}^{-1}$, $740{\rm\,cm}^{-1}$, and  
$840{\rm\,cm}^{-1}$, respectively. 
The third discontinuity of the first derivative labelled ${\rm O}_{3}$ 
can be traced back to the transition ${\rm T}\rightarrow{\rm Q'}$.
It is shifted with respect to ${\rm O}_{1}$ by $E_{X}$, 
the values of the corresponding wavenumbers are 
$850{\rm\,cm}^{-1}$, $970{\rm\,cm}^{-1}$, and  
$1070{\rm\,cm}^{-1}$, respectively. 
Small structures at very low energies 
in parts (b) and (c) of Fig.~\ref{fig:cc1} 
are artefacts due to an incomplete continuation to the real axis. 
The contribution of the cold spot requires a more careful discussion. 
Besides the features ${\rm O}_{1}$, ${\rm O}_{2}$, and ${\rm O}_{3}$,  
it contains a Drude-like component at very low energies. 
Its presence is related to the fact that at this point 
the superconducting gap vanishes. 
Some structures around $60{\rm\,cm^{-1}}$ and $320{\rm\,cm^{-1}}$ 
are due to transitions 
between the maximum of the quasiparticle peak Q located at $-8{\rm\,meV}$  
and its tail ranging to $E=0$,  
and between the tail and the structures R,R', respectively.   
The spectral weight associated with the transition 
${\rm T}\rightarrow {\rm Q'}$ is small  
because only the tail of the quasiparticle peak participates. 
It can be seen in Fig.~\ref{fig:cc1} 
that the spectra for finite $\Gamma$ are similar to those of $\Gamma=0$, 
the onset features, however, are by far less pronounced.
\begin{figure}[htbp]
  \centering 
  \includegraphics[width=0.45\textwidth]{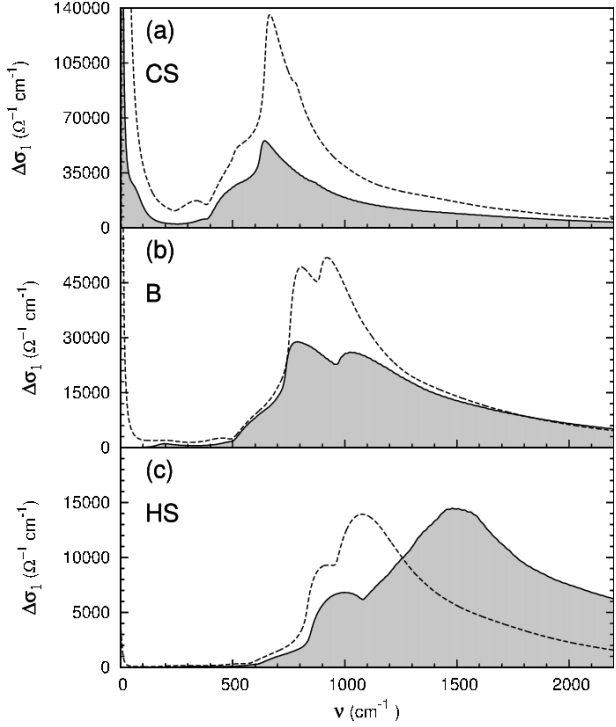}
  \caption{Contributions of the selected $\mathbf{k}$-points 
    to the real part of the optical conductivity from Fig.~\ref{fig:cc1}
 (solid lines)
    compared with those obtained using the values of the input parameters  
    of Ref.~\onlinecite{eschrig03} (dashed lines). } 
  \label{fig:cc2}
\end{figure}

For reference, we compare in Fig.~\ref{fig:cc2} 
the results presented above  
with those obtained using the values of the input parameters 
of Ref.~\onlinecite{eschrig03}  
(the values of $a$, $d$ and $\delta$ have not been changed).   
It can be seen that the frequencies of the singularities 
are fairly close to each other, except for ${\rm O}_{3}$.
The discrepancy can be traced back 
to the difference in the location of the Fermi surface. 
The differences in the absolute values 
reflect those of the dispersion relation and of the coupling constant. 

The contribution of the hot spot  
shown in the bottom panels of Figs.~\ref{fig:cc1} and \ref{fig:cc2}
can be compared with $\sigma_{1}(\omega)$ 
resulting from the model of ACS, 
shown in Fig.~2 of Ref.~\onlinecite{abanov01A}. 
The latter exhibits 
a sharp onset feature at $2\Delta_{0}+\hbar\omega_{0}$, 
a maximum around $5\Delta_{0}$,  and 
a gradual decrease at higher frequencies. 
Our spectra display all these features. 
The onset is labelled as ${\rm O}_{2}$.     
The maximum is located at about $1500{\rm\,cm^{-1}}$ 
for our values of the input parameters
and at ca $1100{\rm\,cm^{-1}}$ for those of Ref.~\onlinecite{eschrig03},  
its frequency strongly depends on the value of $g$.  
The main differences are: 
(a) In the spectra of Ref.~\onlinecite{abanov01A},  
$\sigma_{1}=0$ for frequencies below the onset.  
In our spectra, 
$\Delta\sigma_{1}$ acquires nonzero values already  
above the first onset frequency 
of $\sim \Delta_{0}+\hbar\omega_{0}$. 
The discrepancy is due to the fact that ACS neglected 
excitations involving nodal quasiparticles. 
(b) In our spectra, the onset ${\rm O}_{2}$ 
is followed by another one at ca $\Delta_{0}+E_{X}+\hbar\omega_{0}$ 
(${\rm O}_{3}$).   
Its absence in the spectra of Ref.~\onlinecite{abanov01A} 
is due to the fact that details of the dispersion relation 
have been neglected. 

In summary,  the contribution of a $\mathbf{k}$-point 
to $\sigma_{1}(\omega)$  
exhibits three onset features, 
whose energies are   
${\bar E}_{\mathbf{k}}+\hbar\omega_{0}$, 
ca ${\bar E}_{\mathbf{k}}+\Delta_{0}+\hbar\omega_{0}$, and 
${\bar E}_{\mathbf{k}}+\hbar\omega_{0}+E_{X}$. 

\boldmath 
\subsection{\label{ssec:icd} Real part of the optical conductivity 
and the origin of the onset feature around 
$\omega_{0}$}
\unboldmath

So far we have discussed singularities 
that appear in the contributions to $\sigma_{1}(\omega)$ 
of the individual $\mathbf{k}$-points. 
It is important to find out 
how do these singularities manifest themselves 
in the spectra of the total conductivity. 
Here we shall show that the feature ${\rm O}_{1}$ of Fig.~\ref{fig:cc1}
gives rise to a clear onset of $\sigma_{1}$ 
starting around $\omega_{0}$. 
The singularities labelled as ${\rm O}_{2}$ and ${\rm O}_{3}$ 
cannot be clearly resolved in the spectra themselves,
the former one, however, 
can be uncovered by means of numerical differentiation. 
This will be shown in the next subsection. 

\begin{figure}[htbp]
  \centering
  \includegraphics[width=0.4\textwidth]{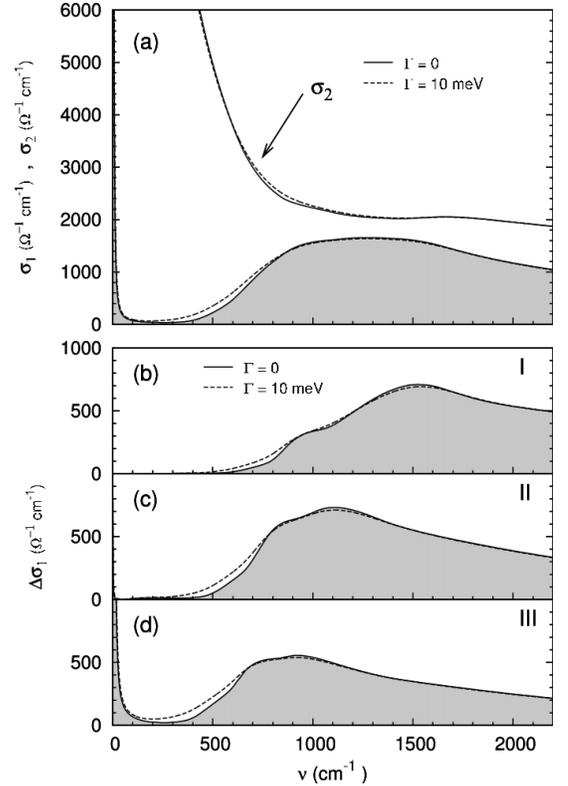}
  \caption{Real and imaginary parts of the optical conductivity (upper panel) 
    and the contributions to $\sigma_{1}(\omega)$
    of the three regions of the irreducible BZ defined in
    Fig.~\ref{fig:zn} (bottom panel). }
  \label{fig:c1}
\end{figure}

\begin{figure}[htbp]
  \centering
   \mbox{\includegraphics[angle=-90,width=0.3\textwidth]{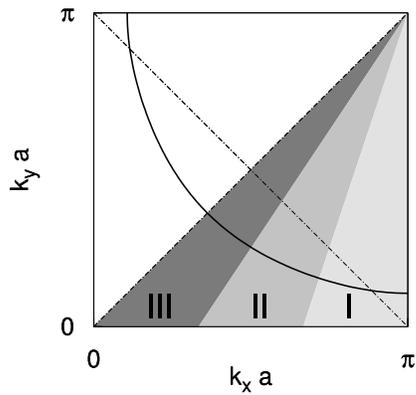}}
  \caption{Three regions of the irreducible BZ
    used in the discussion of the total conductivity. }
  \label{fig:zn}
\end{figure}

\begin{figure}[htbp]
  \centering 
  \includegraphics[width=0.45\textwidth]{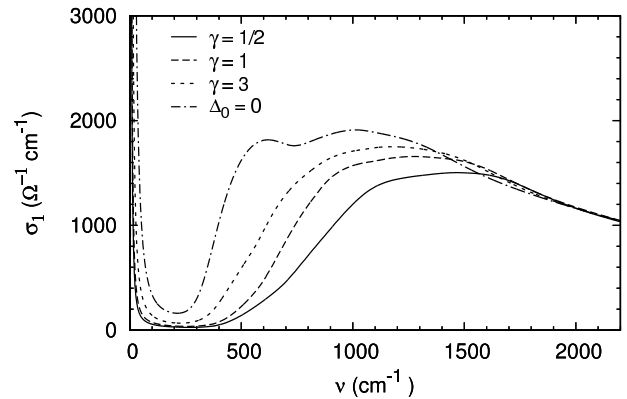}
  \caption{Real part of the optical conductivity for the three forms 
    of the superconducting gap of Eq.~\ref{eq:gpp}. 
    Also shown is the result for $\Delta_{k}=0$. }
  \label{fig:cgp}
\end{figure}

Figure \ref{fig:c1} shows the spectra of $\sigma_{1}(\omega)$ (upper panel) 
and the contributions of the three regions of the irreducible BZ  
defined in Fig.~\ref{fig:zn} (bottom panel).
The imaginary part $\sigma_{2}(\omega)$  
has been obtained by using Eq.~\ref{eq:KKtr}. 
The values of $\omega_{\rm pl}$ and $\omega_{\rm pl,sc}$ 
obtained by using Eq.~\ref{eq:sumr} and Eq.~\ref{eq:omplsc2} are
$17800{\rm\,cm^{-1}}$ and $12100{\rm\,cm^{-1}}$, respectively.  
Interestingly, the magnitudes of the three contributions to $\sigma_{1}(\omega)$
in the infrared are comparable. 
This is a result of a competition between the velocity factor 
in Eqs.~\ref{eq:cond}, \ref{eq:ccond},  
which is higher for the quasiparticles around the BZ diagonal (region III),   
and the density of states at the Fermi surface,   
which is higher in region I. 
It appears that the characteristic singularities  
discussed in the previous subsection can 
still be resolved in the spectra 
of the bottom panel.
For example, the contribution of region I
still contains the onset features ${\rm O}_{1}$, ${\rm O}_{2}$, and 
${\rm O}_{3}$ characteristic of the hot spot. 
In the spectra of the total conductivity, however,
they are approximately averaged out and cannot be clearly resolved, 
except for the first onset feature: 
$\sigma_{1}$ is rather small 
above the narrow Drude peak and below $\omega_{0}$
and it starts to increase 
at $\omega_{0}$ (or, for finite $\Gamma$, around $\omega_{0}$).  
The bottom panel of Fig.~\ref{fig:c1} 
shows that the onset originates in region III.  
It can be attributed to the lowest allowed transitions 
(${\rm Q}\rightarrow{\rm R'}$ and ${\rm R}\rightarrow {\rm Q'}$)
of the region around the nodes.
The shape of the onset can thus be expected to depend   
on the behavior of $\Delta_{\mathbf{k}}$ in the proximity of the nodes. 
In order to explore this dependence, 
we have performed the computations 
for the following three forms of the superconducting gap:  
\begin{equation}
  \Delta_{\mathbf{k}} = \frac{\Delta_0}{2}[\cos(k_xa)-\cos(k_ya)]^\gamma
  \label{eq:gpp}
\end{equation}
with $\gamma=0.5,1,3$ 
(for $\gamma=0.5$ the signs have to be properly adjusted). 
The results are shown in Fig.~\ref{fig:cgp}. 
They can be interpreted as follows: 
the larger (smaller) the value of the exponent $\gamma$, 
the wider (narrower) the region around the BZ diagonal,
where the magnitude of the superconducting gap is small,  
and the larger (smaller) the spectral weight right above $\omega_{0}$ 
originating in this region.  
The increase (decrease) of $\gamma$ thus effectively
results in a red (blue) shift of the onset.

\begin{figure}[htbp]
  \centering
  \includegraphics[width=0.45\textwidth]{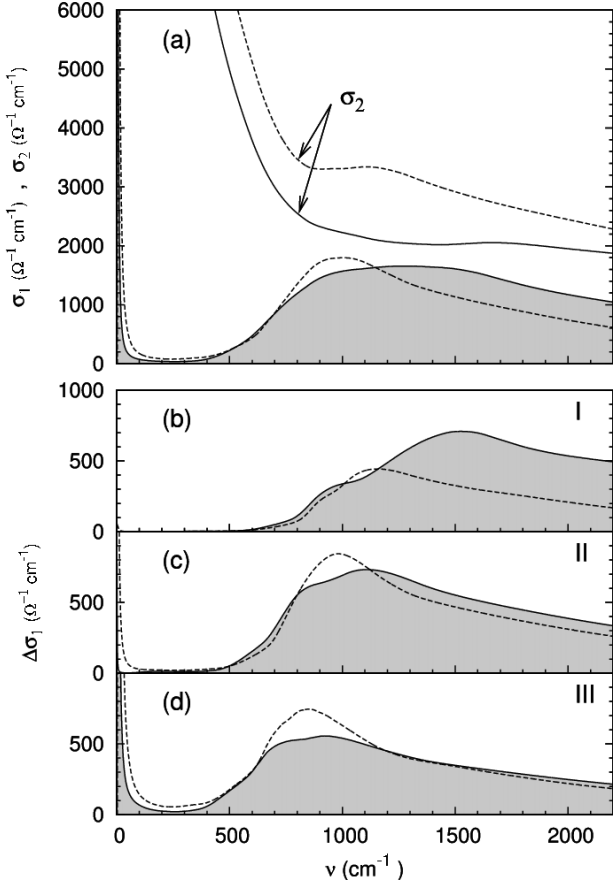}
  \caption{Real and imaginary parts of the optical conductivity (upper panel) 
    and the the three components of $\sigma_{1}(\omega)$ (bottom panel) 
    from Fig.~\ref{fig:c1} (solid lines)
    compared with those obtained using the values of the input parameters  
    from Ref.~\onlinecite{eschrig03} (dashed lines). } 
  \label{fig:c2}
\end{figure}

For reference, we again compare in Fig.~\ref{fig:c2} 
the results presented above  
with those obtained using the values of the input parameters 
of Ref.~\onlinecite{eschrig03};    
the corresponding values of $\omega_{\rm pl}$ and $\omega_{\rm pl,sc}$ 
are $17900{\rm\,cm^{-1}}$ and $12700{\rm\,cm^{-1}}$, respectively.  

\boldmath 
\subsection{\label{ssec:dmW} Parameters of the extended Drude model, 
             effective spectral function $W(\omega)$ and 
             the origin of its main maximum}
\unboldmath

\begin{figure}
\centering
\includegraphics[width=.45\textwidth]{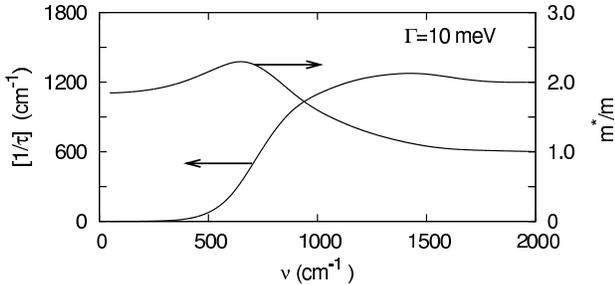}
\caption{Spectra of the scattering rate and the mass enhancement factor  
extracted from those of $\sigma_{1}$ and $\sigma_{2}$ 
shown in Fig.~\ref{fig:c1}.}  
\label{fig:tauandm}
\end{figure}

A common way to discuss the in-plane infrared spectra is 
in terms of the scattering rate $[1/\tau](\omega)$ 
and the mass enhancement factor $[m^{*}/m](\omega)$. 
These quantities are defined by the so called extended Drude model 
\cite{puchkov96,Schulga}: 
\begin{equation}
  \sigma(\omega)=\varepsilon_{0}
  \frac{\omega_{\rm pl}^{2}}{[1/\tau] (\omega)
      -i\omega [m^{*}/m](\omega)}\,.    
  \label{eq:drex}
\end{equation}
Their physical significance relies 
on a close relation between the optical selfenergy, 
\begin{equation} 
\Sigma^{opt}(\omega)=
{1\over 2}\left(
\omega\left[1-{m^{*}\over m}(\omega) \right]
-{i\over \tau}(\omega)
\right) \,,
\label{eq:optself}
\end{equation}
and the (true) quasiparticle selfenergy 
(for a recent discussion see Ref.~\onlinecite{hwang04}).
The calculated spectra of $1/\tau$ and $m^{*}/m$ are shown 
in Fig.~\ref{fig:tauandm}. 
They have been extracted from those of $\sigma_{1}$ and $\sigma_{2}$ 
using the following formulas resulting from Eq.~\ref{eq:drex}:  
\begin{equation}
{1\over \tau}(\omega)=
\omega_{\rm pl}^{2}\varepsilon_{0}
{\sigma_{1}(\omega)\over 
\sigma_{1}^{2}(\omega)+\sigma_{2}^{2}(\omega)}\,,
\label{eq:scrat}
\end{equation}
\begin{equation}
{m^{*}\over m}(\omega)=
{\omega_{\rm pl}^{2}\varepsilon_{0}\over \omega}
{\sigma_{2}(\omega)\over 
\sigma_{1}^{2}(\omega)+\sigma_{2}^{2}(\omega)}\,.
\label{eq:massr}
\end{equation}
The trends in the spectra of $1/\tau$ and $m^{*}/m$
can be easily related to those of $\sigma_{1}$ and $\sigma_{2}$. 
The obvious similarity between the spectra of $1/\tau$ 
and those of $\sigma_{1}(\omega)$, for example, can be interpreted as follows. 
At low  frequencies  
(a) $\sigma_{2}\gg\sigma_{1}$ and 
(b) $\sigma_{2}$ is dominated by the singular term, 
$\varepsilon_{0}\,\omega_{\rm pl,sc}^{2}/\omega$ (see Fig.~\ref{fig:c1}). 
As a consequence,
$[1/\tau](\omega)$ 
is approximately proportional to $\omega^{2}\sigma_{1}(\omega)$
(see Eq.~\ref{eq:scrat}). 
This explains the similarity. 
The latter allows us to consider derivatives of $[1/\tau](\omega)$
instead of those of $\sigma_{1}(\omega)$ 
when searching for the singularities. 

Recently, it has become popular 
\cite{carbotte99,schachinger00,singley01,tu02,wang03,dordevic04}  
to use the function $W(\omega)$ of Eq.~\ref{eq:wom2} 
which is related to $[1/\tau]$ as 
\begin{equation} 
W(\omega)={1\over 2\pi}
{{\rm d}^{2}\over {\rm d}\omega^{2}}
\left[\omega{1\over \tau}(\omega)\right]\,. 
\label{eq:wom3}
\end{equation}
This has been motivated by the theoretical works of 
CSB\cite{carbotte99} and ACS\cite{abanov01A}
discussed in the introduction. 
We recall that according to CSB  
the most pronounced maximum of $W(\omega)$ 
reveals the spectral function of the bosons, 
that are coupled to the charged quasiparticles
(shifted by $\Delta_{0}$, see Eq.~\ref{eq:wom2}). 
Within the resonant mode scenario, the maximum should thus be located  
at $\Omega_{{\rm CSB}}=\Delta_{0}+\hbar\omega_{0}$.
According to the strong-coupling SF-based theory of ACS
$W(\omega)$ should exhibit three singularities at
$2\Delta_{0}+\hbar\omega_{0}$, $4\Delta_{0}$, 
and $2\Delta_{0}+2\hbar\omega_{0}$ (see Fig.~3 of Ref.~\onlinecite{abanov01A}). 
The first singularity, which is the most pronounced one,  
is due to the first onset of $\sigma_{1}(\omega)$  
(see Fig.~2 of Ref.~\onlinecite{abanov01A} and the discussion 
in Subsect.~\ref{ssec:kpt}). 
It consists of a sharp maximum followed by a minimum. 

Here we shall follow the strategy of the preceding subsections 
and attempt to interprete the structures of $W(\omega)$ 
in terms of the quasiparticle spectral functions and 
the corresponding contributions to $\sigma_{1}(\omega)$. 
Our most important finding is that the main maximum of $W(\omega)$
can be associated with the onset feature ${\rm O}_{2}$ 
of parts (a) and (b) of Fig.~\ref{fig:cc1}, 
i.e., with the appearence (above the characteristic frequency of the maximum) 
of final states consisting of 
a quasiparticle from the nodal region, 
a quasiparticle from the antinodal region, 
and the resonance mode. 

\begin{figure}[htbp]
  \centering 
  \includegraphics[width=0.45\textwidth]{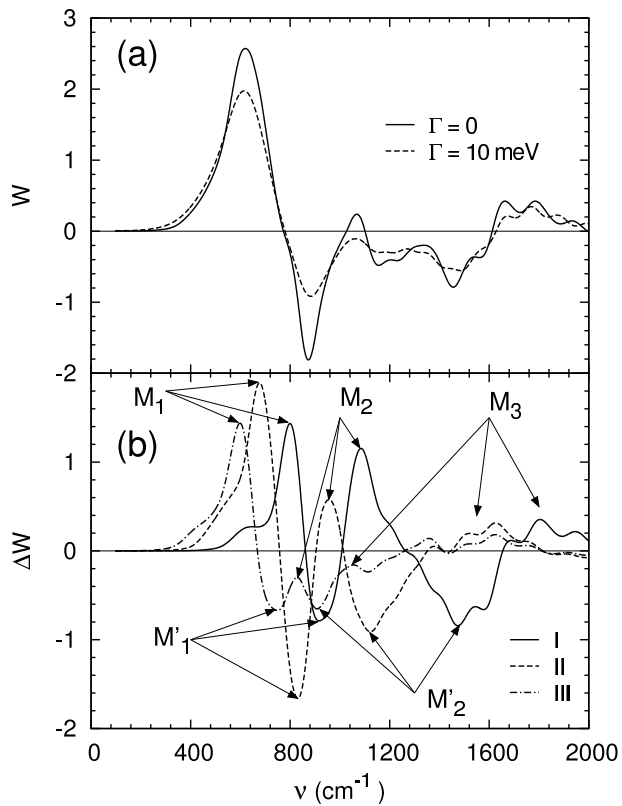}
  \caption{Spectra of the function $W(\omega)$  
    defined by Eq.~\ref{eq:wom2} obtained from the conductivities  
    of Fig.~\ref{fig:c1} (upper panel).  
    Bottom panel:  
     the contributions of the three
    regions defined in Fig.~\ref{fig:zn}.}
  \label{fig:w1}
\end{figure}

Figure \ref{fig:w1} shows our calculated spectra of $W(\omega)$ 
(upper panel) and contributions to $W$ of the three regions 
defined in Fig.~\ref{fig:zn} (bottom panel). The contribution 
of a region is defined by 
\begin{equation}
\Delta W(\omega)=
{\varepsilon_{0}\omega_{pl}^{2}\over 2\pi}
  {{\rm d}^{2}\over {\rm d}\omega^{2}}
\left[\omega
{\Delta\sigma_{1}(\omega)\over 
\sigma_{1}^{2}(\omega)+\sigma_{2}^{2}(\omega)}\right]\,. 
\label{eq:wom4} 
\end{equation}
The spectra of the quantities in the square brackets 
of Eqs.~\ref{eq:wom3} and \ref{eq:wom4} 
have been smoothed 
before taking the second derivatives, 
a typical scale of the smoothing 
being $3{\rm\,meV}$. 
First we discuss the spectra of the bottom panel. 
Each contribution exhibits three maxima ($M_{1}$, $M_{2}$, $M_{3}$) 
separated by two minima ($M'_{1}$, $M'_{2}$). 
The corresponding frequencies are  
$800{\rm\,cm^{-1}}$, $1090{\rm\,cm^{-1}}$, $1800{\rm\,cm^{-1}}$, 
$910{\rm\,cm^{-1}}$, $1500{\rm\,cm^{-1}}$ (I), 
$680{\rm\,cm^{-1}}$, $950{\rm\,cm^{-1}}$, $1620{\rm\,cm^{-1}}$, 
$830{\rm\,cm^{-1}}$, $1130{\rm\,cm^{-1}}$ (II), 
$600{\rm\,cm^{-1}}$, $880{\rm\,cm^{-1}}$, $1030{\rm\,cm^{-1}}$, 
$770{\rm\,cm^{-1}}$, $940{\rm\,cm^{-1}}$ (III). 
Spectral features above the frequency of the third maximum 
are weak. 
It can be relatively easily seen that the maxima $M_{1}$ and $M_{2}$ 
correspond to the onset features ${\rm O}_{2}$ and ${\rm O}_{3}$, respectively 
(see Figs.~\ref{fig:cc1} and \ref{fig:c1})
and that the mimimum $M'_{2}$ is related 
to the maximum of $\Delta\sigma_{1}(\omega)$ 
in mid-infrared (see Fig.~\ref{fig:c1}). 
The structures $M'_{1}$ and $M_{3}$ can be interpreted analogously.  
Next we concentrate on the upper panel. 
The spectra of $W(\omega)$ exhibit 
a pronounced maximum at $\sim 620{\rm\,cm^{-1}}$, 
a minimum at  $\sim870{\rm\,cm^{-1}}$, 
weak structures in the range from  $1000{\rm\,cm^{-1}}$ 
to $1400{\rm\,cm^{-1}}$, 
and a minimum around $1500{\rm\,cm^{-1}}$. 
All these features can be interpreted 
in terms of constructive or destructive interference of the three contributions. 
In particular, the main maximum originates 
from the structures $M_{1}$(III) and $M_{1}$(II), the latter 
being partially compensated by $M'_{1}$(III). 
Note the important role played by the negative contribution of the zone III. 
The frequency of the maximum $\Omega_{M}$ (here $620{\rm\,cm^{-1}}$) 
appears to be slightly higher  
than $\omega(M_{1},III)\approx \omega({\rm O}_{2},III)$ 
(here $600{\rm\,cm^{-1}}$). 
The latter frequency is further slightly higher than 
$\omega({\rm O}_{2},{\rm CS})\approx(\Delta_{0}/\hbar)+\omega_{0}$ 
($565{\rm\,cm^{-1}}$).  
Our general conclusion is that $\Omega_{M}$ is somewhat larger 
than $\Omega_{{\rm CSB}}=(\Delta_{0}/\hbar)+\omega_{0}$ but 
much smaller than  
$2(\Delta_{0}/\hbar)+\omega_{0}$. 
The magnitude of the difference, $\Omega_{M}-(\Delta_{0}/\hbar)-\omega_{0}$, 
depends on details, e.g.,  
on the shape of the superconducting gap. 
In our case we have 
$\Omega_{M}-(\Delta_{0}/\hbar)-\omega_{0}=55{\rm\,cm^{-1}}$. 
The conclusion should be valid for any sharp bosonic mode 
which couples the nodal regions to the antinodal ones. 
It can be easily seen that the shape of the maximum 
does not correspond to the neutron peak. 
The FWHM is about $170{\rm\,cm^{-1}}$ and about $210{\rm\,cm^{-1}}$ 
for $\Gamma=0$ and $\Gamma=10{\rm\,meV}$ 
(ca $80{\rm\,cm^{-1}})$, respectively. 
\begin{figure}[htbp]
  \centering 
  \includegraphics[width=0.45\textwidth]{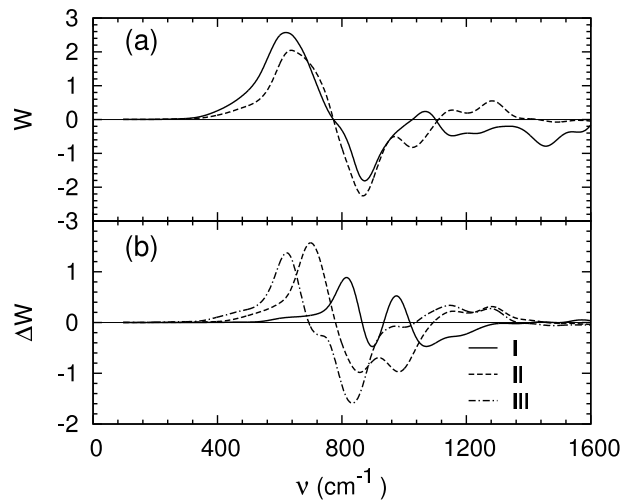}
  \caption{(a) The $\Gamma=0{\rm\,meV}$ spectrum of $W(\omega)$ 
from Fig.~\ref{fig:w1} (solid line) together with the one 
corresponding to the values of the input parameters 
from Ref.~\onlinecite{eschrig03}
(dashed line). (b) 
The same as in Fig.~\ref{fig:w1} (b) 
    but for the values of the input parameters 
    from Ref.~\onlinecite{eschrig03}.} 
  \label{fig:w2}
\end{figure}

Figure~\ref{fig:w2} shows the results 
obtained using the values of the input parameters 
of Ref.~\onlinecite{eschrig03}.   
The main differences between the spectra of the bottom panels
of Fig.~\ref{fig:w1} and Fig.~\ref{fig:w2} are: 
(a) The maxima $M_{1}$ and $M_{2}$ are closer to each other
in Fig.~\ref{fig:w2} than in Fig.~\ref{fig:w1}.  
(b) The maximum $M_{3}$ is located at lower energies in Fig.~\ref{fig:w2} 
than in Fig.~\ref{fig:w1}. 
These differences can be traced back to those 
in the position of the Fermi level and in the value of the coupling constant. 
The frequency of the main maximum in the upper panel of Fig.~\ref{fig:w2}
is approximately the same as that of Fig.~\ref{fig:w1}, 
in agreement with our general conclusion presented above.  
The shape of the maximum and 
the spectral structures at higher energies, however, differ considerably. 

\subsection{\label{ssec:cont} Role of the spin-fluctuation continuum} 

So far we have considered the spin susceptibility 
in the resonant-mode form of Eq.~\ref{eq:MBC}. 
Such an approach may suffice to provide 
an understanding of the spectral features discussed 
in the preceding subsections: 
the onset of $\sigma_{1}(\omega)$ around $\omega_{0}$
and the main maximum of $W(\omega)$. 
It is well known, however, that the spectral weight of the resonant mode 
is much smaller than that of the spin-fluctuation continuum.
The overall shape of the optical response, 
in particular in the mid-infrared region, 
can thus be expected to be determined by the continuum rather than by the mode. 
The inadequacy of the resonant-mode approach to reproduce the mid-infrared data
can indeed be seen 
by comparing the computed scattering rate spectra 
with those deduced from experimental data\cite{vdmarel03,boris04} and 
called ``experimental (scattering rate) spectra" in the following
\cite{epsinfty}.  
In the spectra of Fig.~\ref{fig:tauandm}, 
$1/\tau$ saturates above $1200{\rm\,cm^{-1}}$ 
at a value of about $1200{\rm\,cm^{-1}}$ 
and it decreases at higher frequencies (to be shown below).  
In the experimental spectra, on the other hand,  
$1/\tau$ saturates only above  $\sim 5000{\rm\,cm^{-1}}$,  
at a value of several thousands ${\rm cm^{-1}}$.  

\begin{figure}[htbp]
  \centering 
  \includegraphics[width=0.45\textwidth]{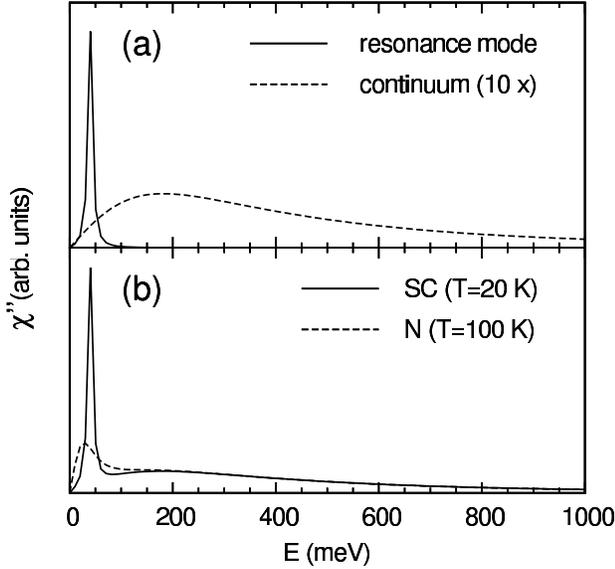}
  \caption{(a) The (${\bf q}$-integrated) spectra 
of the imaginary part $\chi''$ of the spin susceptibility 
corresponding to the resonance mode (solid line) and to the 
continuum term introduced in Eq.~\ref{eq:chiC} (dashed line), respectively. 
(b) Those used as a starting point 
in the computations of Subsec.~\ref{ssec:TD}. 
The solid and the dashed line correspond 
to the superconducting state ($T=20{\rm\,K}$) 
and to the normal state ($T=100{\rm\,K}$), respectively. }  
  \label{fig:mc}
\end{figure}

\begin{figure}[htbp]
  \centering 
  \includegraphics[width=0.5\textwidth]{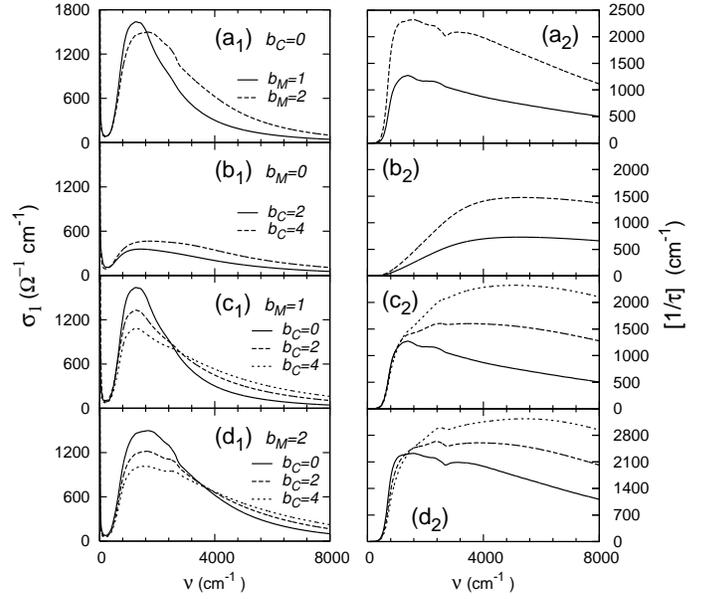}
  \caption{The conductivity and the scattering rate as functions 
of the parameters $b_{\rm M}$ and $b_{\rm C}$ expressing the spectral weights 
of the mode and the continuum, respectively.} 
  \label{fig:sandscrbMbC}
\end{figure}

\begin{figure}[htbp]
  \centering 
  \includegraphics[width=0.5\textwidth]{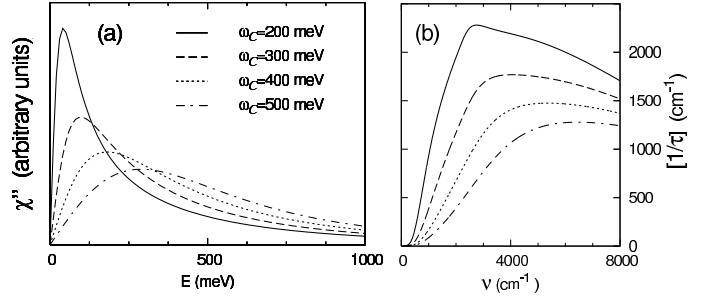}
  \caption{The (${\bf q}$-integrated) imaginary part 
           $\chi''$ of the spin susceptibility (a) 
   and the scattering rate (b) as functions of the parameter $\omega_{C}$
   introduced in the text.} 
  \label{fig:scrandchi}
\end{figure}

Here we report on our attempts to understand the experimental data 
in terms of a model where the spin susceptibility contains, 
in addition to the resonant mode,   
a very broad continuum term: 
\begin{equation}
\chi({\bf q},\omega)=b_{\rm M}\chi_{\rm RM}({\bf q},\omega)+
b_{\rm C}\chi_{\rm C}({\bf q},\omega)\,, 
\label{eq:chiMC}
\end{equation}
where $\chi_{\rm RM}$ is given by Eq.~\ref{eq:MBC} 
with $\Gamma=10{\rm meV}$ in conjunction with 
Eq.~\ref{eq:normalizationchi} 
and 
\begin{equation}
\chi_{\rm C}({\bf q},\omega)=
{1\over 
{1+({\bf q} -{\bf Q})^{2}\,\xi_{\rm C}^{2}}}\,
 {F_{\rm C}\over {\omega_{\rm C}^{2}-\omega^{2}-i\Gamma_{\rm C}\omega}} 
\label{eq:chiC}
\end{equation}
is the continuum component with the same structure as $\chi_{\rm RM}$.
The values of the parameters of $\chi_{\rm C}$ used in our calculations are: 
$\xi_{C}=2{\rm\,\AA}$ 
(i.e., about one half of the value of the lattice parameter),   
$\omega_{\rm C}=400{\rm\, meV}$ (this choice will be motivated below), 
$\Gamma_{\rm C}=1000{\rm\, meV}$, and 
the value of $F_{\rm C}$ is determined by Eq.~\ref{eq:normalizationchi}.
The spectra of both components 
are shown in Fig.~\ref{fig:mc} (a). 
A similar continuum component has been considered 
by Schachinger, Tu, and Carbotte \cite{schachinger03} 
in their discussion of the scattering rate in Bi-2212. 
The parameters $b_{\rm M}$ and $b_{\rm C}$ in Eq.~\ref{eq:chiMC}
express the spectral weights of the two components.
Figure \ref{fig:sandscrbMbC} shows the $b_{\rm M}$- 
and $b_{\rm C}$-dependence of $\sigma_{1}(\omega)$ and $[1/\tau](\omega)$.
The low frequency parts of the spectra shown 
as the solid lines in panels (a$_{1}$) and (a$_{2}$), 
have already been presented above in Figs.~\ref{fig:c1} and \ref{fig:tauandm}. 
Note that in the mid-infrared region  
(a) $\sigma_{1}$ decreases with increasing frequency much faster than 
in the experimental spectra and 
(b) $1/\tau$ also decreases, 
in contrast to the experimental spectra.  
The right panel of Fig.~\ref{fig:sandscrbMbC} 
clearly demonstrates the distinct roles of the two components: 
(i) The mode yields a relatively sharp onset of $1/\tau$
centered at about $\Omega_{\rm M}$,    
followed by a gradual decrease.  
(ii) The continuum leads to an approximately linear increase
followed by a plateau above  $\sim 4000{\rm\,cm^{-1}}$. 
As shown in Fig.~\ref{fig:scrandchi}, 
the onset frequency of the plateau reflects 
the width of $\chi_{\rm C}$, 
which is here determined by $\omega_{\rm C}$ of Eq.~\ref{eq:chiC}.
Our choice of the value of $\omega_{\rm C}$ 
has been motivated by the experimental spectra 
of Refs.~\onlinecite{vdmarel03} and \onlinecite{boris04}.  
(iii) It can be seen that $1/\tau$ is approximately 
a linear function of $b_{M}$ and $b_{C}$. 
For example, $[1/\tau](b_{M}=1,b_{C}=4)$ [dotted line in part (c$_{2}$)]
is approximately equal to $[1/\tau](b_{M}=1,b_{C}=0)$ 
[solid line in part (a$_{2}$)] $+$ 
$[1/\tau](b_{M}=0,b_{C}=4)$ 
[dashed line in part (b$_{2}$)]. 
The value of $b_{M}$ ($b_{C}$) determines the value of $1/\tau$  
above the sharp onset or, 
in other words, the height of the step in $1/\tau$ 
(the difference between the value of $1/\tau$ of the plateau 
and the height of the step). 
(iv) The spectra for $b_{M}\not=0$ 
exhibit an additional structure around $2500{\rm\,cm^{-1}}$.  
It can be shown, along the lines of Subsect.~\ref{ssec:kpt}, 
to be connected to a singularity in the density of the relevant final states 
located approximately at 
\begin{equation}
\hbar\omega_{S}=
\mu-\epsilon_{\Gamma}+\Delta_{0}+\hbar\omega_{0}\,.  
\label{eq:omegaS}
\end{equation}  
The latter is related to the van Hove singularity 
associated with the bottom of the band 
at the $\Gamma$-point. 
The frequency $\omega_{S}$ can thus be used to estimate 
$\mu-\epsilon_{\Gamma}$,  
i.e., the width of the occupied part of the band. 
It is very interesting that the experimental scattering rate spectra, 
both for Y-123 and for Bi-2212,  
display a fine structure around $3000{\rm\,cm^{-1}}$, 
which is similar to that appearing in the computed spectra.  
The similarity leads us to speculate 
that it has the same origin, 
i.e., that it is due to the singularity of the density of states 
discussed above. 
The value of $\mu-\epsilon_{\Gamma}$ of $\sim 0.3{\rm\,eV}$ 
resulting from Eq.~\ref{eq:omegaS}  
is in rough agreement with the photomession values 
of  $0.25{\rm\,eV}$ (Y-123, Ref.~\onlinecite{shen95}) 
and $0.40{\rm\,eV}$ (Bi-2212, Ref.~\onlinecite{damascelli03}). 
According to the proposed interpretation, 
the structure should become less pronounced above $T_{c}$ 
because of the disappearence of the sharp mode. 
In the experimental spectra, however, 
it seems to be only weakly temperature dependent. 
At present we are not aware of any explanation of this discrepancy. 

It can be seen  
that the experimental spectra can be approximately reproduced 
only by taking a value of $b_{\rm M}$ of around $1.0$ 
and a considerably higher value of $b_{\rm C}$. 
For $b_{\rm M}=1.0$ and $b_{\rm C}=4.0$, e.g., 
the calculated spectra are in reasonable agreement 
with those of optimally doped Y-123 from Ref.~\onlinecite{boris04}.
In order to reproduce the spectra of optimally doped Bi-2212 
from Ref.~\onlinecite{vdmarel03}
a higher value of $b_{\rm C}$ (ca 10) has to be taken. 
Let us note that 
the values of the ratio $b_{\rm C}/b_{\rm M}$ under consideration  
are still smaller than those suggested 
by the neutron scattering experiments \cite{fong00,fong99}. 

\subsection{\label{ssec:TD} Some aspects of the temperature dependence 
of the spectra} 

In Subsubsect.~\ref{sssec:TDsig} 
we present and discuss a computed temperature dependence of the spectra 
characteristic of the spin-fluctuation scenario, 
which may be helpful for understanding the experimental data. 
In Subsubsect.~\ref{sssec:TDSW} 
we then focus on the temperature dependence 
of the intraband spectral weight.  

\subsubsection{\label{sssec:TDsig} Temperature dependence 
of the infrared spectra characteristic of the spin-fluctuation scenario} 

\begin{figure}[htbp]
  \centering 
  \includegraphics[width=0.45\textwidth]{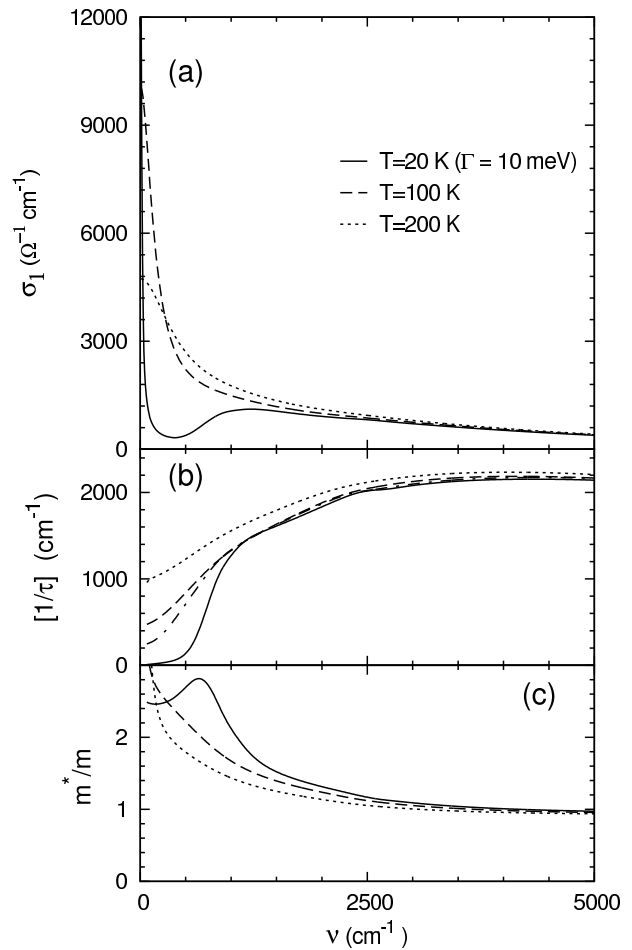}
  \caption{Calculated temperature dependencies of the conductivity, 
the scattering rate, and the mass enhancement factor. 
The dashed-dotted line in part (b) corresponds to an unrealistic case 
of the normal state, $T=100{\rm\, K}$, and 
the spin susceptibility containing the sharp mode.} 
  \label{fig:TDsscrm}
\end{figure}

Figure \ref{fig:TDsscrm} shows the spectra of $\sigma_{1}$, $1/\tau$ 
and $m^{*}/m$ computed for three temperatures: 
$T=20{\rm\,K}$ (superconducting state), $T=100{\rm\,K}$, and $T=200{\rm\,K}$ 
(normal state). 
For each temperature, the value of the chemical potential was adjusted 
to yield the total number of electrons of 0.76 per unit cell: 
$\mu=-355.5{\rm\,meV}$, $-347.5{\rm\,meV}$, and $-345.7{\rm\,meV}$ 
for $T=20{\rm\, K}$, $100{\rm\,K}$, and $200{\rm\,K}$, respectively. 
The values of the other input parameters for the normal state  
are the same as those for the superconducting state given in Table 1,  
except for 
$\Delta_{0}=0$, 
$\Gamma=70{\rm\,meV}$ (the mode is assumed to be overdamped),  
and $\xi=6{\rm\,\AA}$. 
Both for the superconducting and for the normal state 
we furher include the continuum contribution to $\chi$ 
in the form introduced in Subsect.~\ref{ssec:cont}, 
with four times larger spectral weight 
than that of the mode ($b_{\rm C}/b_{\rm M}=4$).  
The $20{\rm\, K}$ and $100{\rm\,K}$ spectra of $\chi''$ 
are shown in Fig.~\ref{fig:mc} (b). 
The calculated values of $\omega_{\rm pl}$ and $\omega_{\rm pl,sc}$
are $17500{\rm\,cm^{-1}}$ and $10500{\rm\,cm^{-1}}$, respectively.

It can be seen that the model is capable of reproducing 
not only the main characteristics of the low temperature infrared spectra  
but also, at least qualitatively,  
the differences between the normal and the superconducting state. 
Our discussion will be limited to the following issues:  

(i) {\it The frequency dependence of }
$\Delta\sigma_{1}=
\sigma_{1}(100{\rm\,K})-\sigma_{1}(20{\rm\,K})$ {\it in mid-infrared}. 
In both the experimental (see, e.g., Ref.~\onlinecite{boris04})
and the computed spectra 
(see Fig.~\ref{fig:TDsscrm}), 
$\Delta\sigma_{1}$ is 
a positive and (approximately) decreasing function of frequency. 
In the experimental spectra  
the decrease is very rapid below $\sim 1000{\rm\,cm^{-1}}$,  
i.e., in the region of the spectral gap. 
Above $1000{\rm\,cm^{-1}}$, 
the superconducting and the normal state spectra are close to each other
and $\Delta\sigma_{1}$ is very small. 
In the computed spectra, on the other hand, the decrease of $\Delta\sigma_{1}$ 
is relatively gradual. 
The discrepancy is very probably due to the fact 
that our approach is not self-consistent.  
We believe that it would vanish or diminish
if a selfconsistent version of Eq.~\ref{eq:Selfen3}
(i.e., $G_{0}$ replaced with $G$) was used. 

(ii) {\it The onset of $1/\tau$}.    
The $100{\rm\,K}$ spectrum does not display 
the sharp onset around $700{\rm\,cm^{-1}}$ discussed above. 
Nevertheless, at low frequencies it deviates slightly 
from the straight line that can be used to approximate $[1/\tau](\omega)$
in the region between ca $1000{\rm\,cm^{-1}}$ and $2500{\rm\,cm^{-1}}$, 
an effect similar to the one 
of some experimental spectra \cite{tu02,bernhard02,hwang04,boris04}. 
Obviously, this kind of deviation does not simply imply 
the presence of a sharp mode. 
For comparison we also show in Fig.~\ref{fig:TDsscrm} (b) 
the spectrum corresponding to an unrealistic case 
of the normal state, $T=100{\rm\, K}$, and 
the spin susceptibility containing the sharp mode ($\Gamma=10{\rm\, meV}$, 
$\xi=9{\rm\,\AA}$). 
Here the onset feature is more pronounced. 

(iii) {\it The fine structure around} $2500{\rm\,cm^{-1}}$. 
It is sharper in the superconducting state than in the normal state. 
This seems not to apply to the similar structure 
of the experimental scattering rate spectra. 

(iv) {\it The onset of the plateau of the scattering rate spectra}.  
With increasing temperature, it shifts towards lower frequencies. 
This is consistent with the experimental data of Ref.~\onlinecite{vdmarel03}. 

(v) {\it The normal-state $m^{*}/m$ spectra --- low-frequency asymptotics}. 
Below  $\sim 300{\rm\,cm^{-1}}$, $m^{*}/m$ 
rapidly increases with decreasing frequency. 
This is an artefact of our computational approach 
which has an interesting physical background.
We calculate $\sigma_{2}(\omega)$ systematically using Eq.~\ref{eq:KKtr} 
and $\langle K\rangle $ by taking the average 
of the expression on the right hand side of  Eq.~\ref{eq:diamt}.
The value of $\langle K\rangle$ is very slightly 
(by about $2{\rm\,\%}$) larger than 
the one given by the sum rule of Eq.~\ref{eq:sumr}. 
As a consequence, $\sigma(\omega)$ contains 
a small unphysical singular part,  
which is responsible for the behaviour of $m^{*}/m$ 
at low frequencies.  
We have checked that the problem is not due to a limited precision 
of our numerical calculations. 
Instead, it seems to be caused 
by the neglect of vertex corrections in Eq.~\ref{eq:pij}.  
The latter have been recently suggested to influence the response functions 
dramatically \cite{drew04,millis04}. 
For the present values of the parameters, their role,  
at least as the integrated spectral weight is concerned, 
appears to be only minor. 
They can be expected, however, 
to become more pronounced for higher values of $g$. 
Indeed, we have found, that 
the discrepancy between 
the results of Eqs.~\ref{eq:diamt} and Eq.~\ref{eq:sumr} 
mentioned above 
increases considerably with increasing $g$. 
Note finally that for $\omega>500{\rm\,cm^{-1}}$ 
the spectra of $m^{*}/m$ are almost unaffected 
by the spurious component of $\sigma$. 

(vi) {\it High-frequency asymptotics of $m^{*}/m$}. 
At higher frequencies,  
$[m^{*}/m](\omega,T=20{\rm\,K})>[m^{*}/m](\omega,T=100{\rm\,K})$.
For the present values of the parameters,  
the two spectra converge only slowly to their common limit of 1. 
At $\omega=5000{\rm\,cm^{-1}}$, 
the magnitude of the difference is still approximately 0.02. 

\subsubsection{\label{sssec:TDSW} Temperature dependence  
of the total intraband spectral weight} 

The changes of the total intraband spectral weight, 
$I_{\rm O}=\int_{0}^{\infty}{\rm d}{\omega}\,\sigma_{1}(\omega)$, 
are of particular interest, 
because this quantity is approximately proportional 
to the effective kinetic energy of the charged quasiparticles.    
The latter is defined by  
\begin{equation}
{\rm K.E.}=
\sum_{\mathbf{k}}
\varepsilon_{\mathbf{k}}n_{\mathbf{k}}\,,\,
n_{\mathbf{k}}=\sum_{\alpha}
\langle c^+_{\mathbf{k}\,\alpha}c_{\mathbf{k}\,\alpha}\rangle\,  
\label{eq:Efke}
\end{equation}
with an obvious extension to multilayer compounds.
In the case of the nearest-neighbor tight-binding dispersion relation,
the sum rule of Eq.~\ref{eq:sumr} reduces to
\begin{equation}
I_{\rm O}=
-{\pi\over 2}
{e^{2}\over d\,\hbar^{2}}{1\over 2}\times {\rm K.E.}/{\rm unit\,cell}\,.      
\label{eq:sumrNNTB}
\end{equation}
For very small values --- relative to that of $t$ --- of 
further-neighbor hopping parameters, 
Eq.~\ref{eq:sumrNNTB} can be expected to hold approximately.
The values of the latter parameters 
characteristic of the real high-$T_{\rm c}$ materials 
are smaller yet comparable with $t$.  
It is thus not immediately clear, to what extent Eq.~\ref{eq:sumrNNTB}
applies.  
In order to obtain an insight,
we first concentrate on the trivial case 
of noninteracting quasiparticles ($g=0$). 
In the remaining part of this subsubsection we shall 
discuss the TD of $I_{\rm O}$ and related quantities,  
resulting from the spin-fermion model used above. 

\begin{figure}[htbp]
  \centering 
  \includegraphics[width=0.45\textwidth]{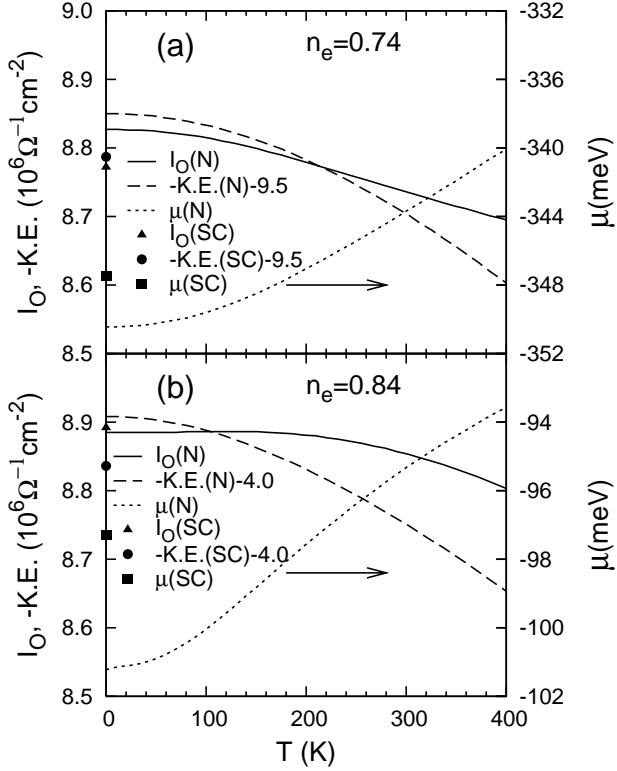}
  \caption{Temperature dependencies 
of the total intraband spectral weight $I_{O}$, 
the effective kinetic energy per unit cell multiplied by 
$-2d\hbar^{2}/(\pi e^{2})$ (labelled as $-{\rm K.E.}$),         
and the chemical potential
for noninteracting quasiparticles. 
Parts (a) and (b) correspond 
to the dispersion relations of Eq.~\ref{eq:dsp} 
and of Ref.~\onlinecite{eschrig03}, respectively. 
For each temperature, 
the value of the chemical potential was adjusted 
to yield the total number of electrons per unit cell 
of 0.74 (a) and 0.84 (b).  
The symbols correspond to the BCS ansatz with 
the superconducting gap given by Eq.~\ref{eq:gap}.}  
  \label{fig:TDBCS}
\end{figure}

Figure \ref{fig:TDBCS} shows the computed temperature dependencies 
of $\mu$,  ${\rm K.E.}/{\rm unit\,cell}$, and $I_{\rm O}$  
for the dispersion relation of Eq.~\ref{eq:dsp} 
and of Ref.~\onlinecite{eschrig03}. 
It can be seen that  
(i) The chemical potential increases monotonically 
with increasing temperature. 
(ii) The kinetic energy increases with increasing temperature, 
approximately following a parabola. 
(iii) In part (a), $I_{O}$ is approximately equal to 
one half of  
$-2d\hbar^{2}/(\pi e^{2}) {\rm K.E.}$,         
in agreement with Eq.~\ref{eq:sumrNNTB}, and  
the TD of $I_{\rm O}$ also tracks the one of ${\rm K.E.}$ 
In part (b), the TD is more complex: 
$I_{\rm O}$ is approximately $T$-independent for $T<200{\rm\,K}$ and 
it decreases at higher temperatures. 
Note that the only generic trend is that of the point (ii). 
The other aspects depend very much on details of the dispersion relation. 
Even for the simple $t-t'$ dispersion --- with  
the present values of the hopping parameters --- both 
$\mu$ and $I_{O}$ can either increase/decrease or decrease/increase
depending on the location of the Fermi surface with respect 
to the Van-Hove singularity associated with the $X$-point. 
The first (second) variant 
occurs for the Fermi surface above (below) the singularity. 
Next we discuss changes of the three quantities 
when going from the normal 
to the superconducting state.  
(i) The chemical potential increases in both cases. 
This can be explained as follows: 
in the superconducting state 
the region of the BZ around the Van-Hove singularity 
becomes partially unoccupied and the chemical potential has to increase 
in order that the particle number would be conserved. 
(ii) The kinetic energy increases, 
a characteristic feature of the BCS theory.   
(iii) In part (a) the spectral weight $I_{\rm O}$ behaves similarly as   
$-{\rm K.E}$, 
in agreement with Eq.~\ref{eq:sumrNNTB}.  
In part (b), however, $I_{\rm O}$ very slightly increases, 
contrary to what one would expect based on Eq.~\ref{eq:sumrNNTB}. 
Again, only the trend of the point (ii) is a generic one. 

\begin{figure}[htbp]
  \centering 
  \includegraphics[width=0.45\textwidth]{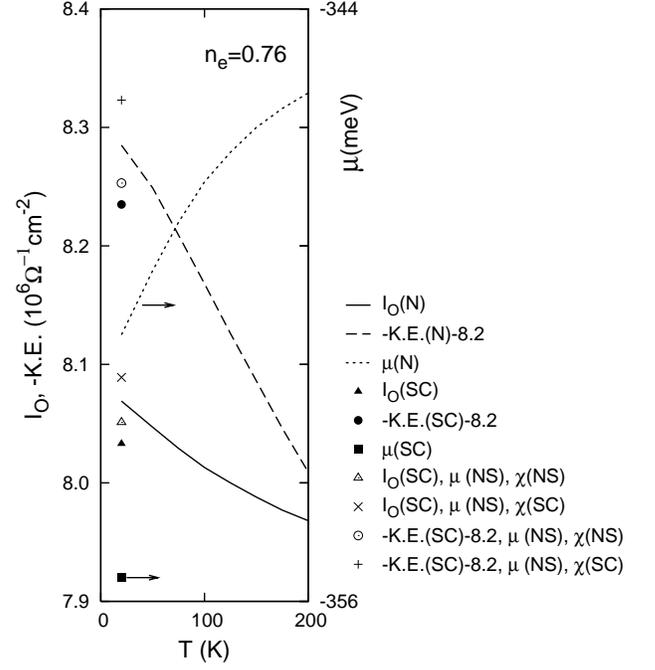}
  \caption{The same as in Fig.~\ref{fig:TDBCS} 
but for quasiparticles coupled to the spin fluctuations.  
The lines correspond to the normal state 
and the solid symbols to the superconducting state. 
For each value of $T$ 
the value of $\mu$ was adjusted 
to yield the total number of electrons per unit cell of 0.76. 
The open symbols (the symbol $\times$ and the star) 
correspond to an unrealistic case 
of the superconducting state, 
$\mu=\mu({\rm normal\,state},T=20{\rm\, K})$, 
$\chi({\bf q},\omega)=\chi({\rm normal\,state},T=20{\rm\, K},{\bf q},\omega)$
[$\chi({\bf q},\omega)=\chi({\rm superconducting\,state},
T=20{\rm\, K},{\bf q},\omega)$].}
  \label{fig:TDSF}
\end{figure}

Figure \ref{fig:TDSF} shows the computed temperature dependencies 
of $\mu$,  ${\rm K.E.}/{\rm unit\,cell}$, and $I_{\rm O}$  
for quasiparticles coupled to the spin fluctuations. 
The dispersion relation of Eq.~\ref{eq:dsp}  
and the model spin susceptibility  
introduced in the preceding subsection and subsubsection 
have been used. 
The normal-state trends are similar to those of Fig.~\ref{fig:TDBCS} (a):
$\mu$ and ${\rm K.E.}$ increase and $I_{\rm O}$ decreases. 
Both the quantities ${\rm K.E.}$, $I_{\rm O}$,  
and their changes, $\Delta I_{O}=I_{O}(200{\rm\,K})-I_{O}(20{\rm\,K})$ and 
$\Delta {\rm K.E.}={\rm K.E.}(200{\rm\,K})-{\rm K.E.}(20{\rm\,K})$, 
approximately fulfill Eq.~\ref{eq:sumrNNTB}. 
The changes occuring upon entering the superconducting state 
(at $T=20{\rm\, K}$) are the following:
$\mu$ decreases by $\sim 5{\rm\,meV}$, 
${\rm K.E.}$ increases, and $I_{\rm O}$ decreases.  
The magnitudes of the changes of ${\rm K.E.}$ and $I_{\rm O}$  
are surprisingly similar to those of Fig.~\ref{fig:TDBCS}, 
$\Delta I_{\rm O}=I_{\rm O}({\rm SC},T=20{\rm\,K})-
I_{\rm O}({\rm N},T=20{\rm\,K})=
-0.04\cdot10^{6}\,\Omega^{-1}{\rm cm^{-2}}$.   
Next we propose an interpretation 
of these effects associated with the superconducting transition. 
For the present purposes and within the context of our approach 
the latter transition can be thought of 
as consisting of three stages: 
(i) Opening of the superconducting gap 
without relaxing the chemical potential 
and the spin susceptibility. 
(ii) ``Improving" (sharpening) of the quasiparticles 
due to the superconductivity related change of the spin susceptibility,
in particular,  due to the formation of the spin gap 
(which is, within our naive model, achieved by the narrowing 
of the low frequency component of $\chi$), 
still without relaxing $\mu$.  
This is the feedback effect of the spin fluctuations on the quasiparticles 
which has been already considered by several authors 
\cite{norman00,norman02,yanase04}.
(iii) Change of the momentum distribution function $n_{\mathbf{k}}$
connected with the change of $\mu$ which is reinforced by the requirement 
of particle-number conservation. 
The first stage results in an increase of ${\rm K.E.}$ 
and a related decrease of $I_{\rm O}$, 
see the open symbols in Fig.~\ref{fig:TDSF}. 
These are the same effects as in Fig.~\ref{fig:TDBCS} (a) 
and their explanation is trivial. 
The second stage is associated with a decrease of ${\rm K.E.}$ and 
an increase of $I_{\rm O}$, 
see the symbols $+$ and $\times$ in Fig.~\ref{fig:TDSF}. 
These effects are basically due to the fact that the superconducting state is,
as regards the momentum distribution function, more Fermi-liquid like 
than the normal state \cite{norman02,yanase04}.   
If it were only for the two stages, (i) and (ii), 
the ${\rm K.E.}$ ($I_{\rm O}$) would---for 
the present values of the input parameters---decrease (increase) 
when going from the normal to the superconducting state. 
The third stage, however, can reverse the result. 
This is what happens for the present values of the input parameters.
A lesson one can learn from the example of this subsubsection 
is that the changes of ${\rm K.E.}$ and $I_{\rm O}$ 
associated with the superconducting transition are determined 
by a delicate competition of the three ingredients: 
the opening of the superconducting gap, 
the feedback effect of the spin fluctuations on the quasiparticles,   
and the chemical potential shift, 
which may be large because of an inherent electron-hole asymmetry. 
So far we have considered changes   
occuring at a fixed temperature of $20{\rm\,K}$. 
These are not directly accessible to experimental investigations 
and can only be obtained 
by a careful elimination of the normal state trend.  
The quantity 
$\Delta' I_{\rm O}=I_{\rm O}({\rm SC},T=20{\rm\,K})-
I_{\rm O}({\rm N},T=100{\rm\,K})$, however,  
can be measured.  
Our calculations yield a small and positive value of $\Delta' I_{O}$ of  
$0.015\cdot10^{6}\,\Omega^{-1}{\rm cm^{-2}}$.     

\section{\label{sec:cnc} Summary and conclusions}

In Subsects.~\ref{ssec:akr}-\ref{ssec:dmW} we have analyzed 
the structures of the superconducting state (SCS) conductivity 
reflecting the coupling of the charge quasiparticles to the resonance mode:   
the onset of the real part $\sigma_{1}(\omega)$ of the conductivity 
starting around the frequency $\omega_{0}$ of the mode 
and the maximum of the function $W(\omega)$ 
centered approximately at $\omega=\omega_{0}+\Delta_{0}/\hbar$, 
where $\Delta_{0}$ is the maximum value of the superconducting gap. 
Our analysis provides a clear interpretation of both features. 
It is based on a division of the Brillouin into three parts: 
the nodal, the region, and the intermediate. 
Both $\sigma_{1}(\omega)$ and $W(\omega)$ can be expressed 
as the sums of contributions of the three parts,  
whose structures can be easily related 
to those of the quasiparticle spectral function $A(\mathbf{k},\omega)$.  
This gives us a possibility to understand also the features 
of $\sigma_{1}(\omega)$ and $W(\omega)$ in terms of $A(\mathbf{k},\omega)$.  
We have shown that the onset is due to the appearance above $\omega_{0}$ 
of low energy excitations of the nodal region, 
consisting of two nodal quasiparticles and the mode. 
For the conventional shape of the $d$-gap the onset is very gradual  
but it becomes steeper with increasing broadening of the nodes.  
The maximum of $W(\omega)$ has been shown to be due 
to the appearance above $\omega_{CSB}=\omega_{0}+\Delta_{0}/\hbar$ 
of excitations of the nodal region,
consisting of a nodal particle, an antinodal particle and the mode.  
Due to the Brillouin zone averaging, 
in particular,  due a considerable contribution of the intermediate region, 
the maximum of $W(\omega)$ appears at a frequency $\Omega_{M}$
which is slightly higher  
but considerably lower than $\omega_{0}+(2\Delta_{0}/\hbar)$.  
This result of our analysis confirms, to some extent,  
the claim by Carbotte, Schachinger, and Basov, 
that the frequency of the maximum is approximately equal to $\Omega_{CSB}$. 
The width of the maximum, however,   
is determined not only by the one of the mode 
but also by the averaging. 
For zero width of the mode, the FWHM of the maximum 
is ca $160{\rm\,cm^{-1}}$. 

In the light of the present analysis, 
the interpretation of the onset feature 
in the SCS infrared spectra 
in terms of the neutron resonance 
remains to be very plausible. 
In fact, we are not aware of any other comparable explanation.
An encouraging finding is that the values of the coupling constant 
to be used in conjunction with the properly normalized spin susceptibility 
are of the same order of magnitude as those of $U$ 
expected to be relevant for the HTCS. 

In Subsect.~\ref{ssec:cont}
we have studied the role of the spin-fluctuation continuum. 
The computed spectra of the scattering rate $1/\tau$ 
exhibit a sharp step (onset feature) centered around $\Omega_{M}$, 
a region of an approximately linear increase of $1/\tau$,   
and a plateau in the mid-infrared. 
The height of the step is determined 
by the spectral weight $I_{R}$ of the mode, 
the difference between the value of $1/\tau$ of the plateau
and the height of the step by the spetral weight $I_{C}$ of the continuum, 
and the onset frequency of the plateau by the width of the continuum.
The experimental scattering rate spectra 
can be reproduced using values of $I_{C}/I_{M}$ of 5-10 and 
those of the width of several hundreds meV. 
The computed spectra of $1/\tau$ display, in addition, 
an interesting weak structure 
located approximately at $\mu-\epsilon_{\Gamma}+\Delta_{0}+\hbar\omega_{0}$
($\mu$ is the chemical potential and $\epsilon_{\Gamma}$ the quasiparticle 
energy of the $\Gamma$-point).  
The presence of this structure in the experimental spectra 
would allow one to estimate the width of the occupied part of the band.   

Finally, in Subsect.~\ref{ssec:TD}
we have concentrated on the temperature dependence of the spectra. 
The main differences
between the SCS and the normal state (NS) experimental data, 
in particular, 
the spectral gap, 
the tail of $\sigma_{1}(\omega,{\rm NS})-\sigma_{1}(\omega,{\rm SCS})$ 
ranging to very high frequencies, 
and the shift of the scattering rate plateau towards lower frequencies 
with increasing temperature are reasonably well reproduced. 
Some of the remaining discrepancies are very likely to be related 
to the absence of self-consistency and vertex corrections. 
We have further investigated the temperature dependence 
of the effective kinetic energy ${\rm K.E.}$ 
and of the intraband spectral weight $I_{\rm O}$,  
first for the BCS case  
and second for the present version of spin-fermion model. 
In the BCS case, ${\rm K.E.}$ increases 
both with increasing temperature in the NS  
and when going from the NS to the SCS. 
The spectral weight behaves similarly as $-{\rm K.E.}$
only for the simpler of the two (realistic) dispersion relations used 
(the ``$t-t'$" dispersion relation)
and for the Fermi surface above the Van-Hove singularity.
For the other dispersion relation 
even the sign of the change of $I_{\rm O}$ 
between the NS and the SCS 
differs from the one of $-{\rm K.E.}$.  
The estimates of the (in-plane) kinetic energy changes 
based on optical data should thus be taken with caution. 
The temperature dependencies of ${\rm K.E.}$ and $I_{\rm O}$ 
calculated for the case of the quasiparticles coupled to the spin fluctuations 
are, for the present values of the input parameters,  
similar to those of the BCS, $t-t'$ case.  
The physics underlying the differences between the NS and the SCS, 
however, is more complicated. 
It involves three different mechanisms causing 
changes of the occupation numbers $n_{\mathbf k}$:  
the opening of the gap 
[leading to an increase (decrease) of ${\rm K.E.}$ ($I_{\rm O}$)],
a sharpening of the quasiparticles 
in the SCS associated with the formation of the spin gap
(resulting in an opposite trend), 
and a chemical potential shift
(leading here to the same trend as the opening of the gap). 
While the mechanisms are certainly generic, 
the outcome of their competition, 
may depend on details.  

\acknowledgments
This work was supported by the Ministry of Education of CR (MSM0021622410). 
D.~M.~thanks the AvH foundation for support during 
a short stay at MPI Stuttgart. 
Discussions with A.~Dubroka, A.~V.~Boris, N.~N.~Kovaleva, and B.~Keimer   
are gratefully acknowledged.


\begin{thebibliography}{65}
\expandafter\ifx\csname natexlab\endcsname\relax\def\natexlab#1{#1}\fi
\expandafter\ifx\csname bibnamefont\endcsname\relax
  \def\bibnamefont#1{#1}\fi
\expandafter\ifx\csname bibfnamefont\endcsname\relax
  \def\bibfnamefont#1{#1}\fi
\expandafter\ifx\csname citenamefont\endcsname\relax
  \def\citenamefont#1{#1}\fi
\expandafter\ifx\csname url\endcsname\relax
  \def\url#1{\texttt{#1}}\fi
\expandafter\ifx\csname urlprefix\endcsname\relax\def\urlprefix{URL }\fi
\providecommand{\bibinfo}[2]{#2}
\providecommand{\eprint}[2][]{\url{#2}}

\bibitem[{\citenamefont{Norman and Pepin}(2003)}]{norman03}
\bibinfo{author}{\bibfnamefont{M.~R.} \bibnamefont{Norman}} \bibnamefont{and}
  \bibinfo{author}{\bibfnamefont{C.}~\bibnamefont{Pepin}},
  \bibinfo{journal}{Rep. Prog. Phys.} \textbf{\bibinfo{volume}{66}},
  \bibinfo{pages}{1547} (\bibinfo{year}{2003}).

\bibitem[{\citenamefont{Abanov et~al.}(2003)\citenamefont{Abanov, Chubukov, and
  Schmalian}}]{abanov03}
\bibinfo{author}{\bibfnamefont{A.}~\bibnamefont{Abanov}},
  \bibinfo{author}{\bibfnamefont{A.~V.} \bibnamefont{Chubukov}},
  \bibnamefont{and}
  \bibinfo{author}{\bibfnamefont{J.}~\bibnamefont{Schmalian}},
  \bibinfo{journal}{Adv in Phys.} \textbf{\bibinfo{volume}{52}},
  \bibinfo{pages}{119} (\bibinfo{year}{2003}).

\bibitem[{\citenamefont{Bickers and Scalapino}(1989)}]{bickers89}
\bibinfo{author}{\bibfnamefont{N.~E.} \bibnamefont{Bickers}} \bibnamefont{and}
  \bibinfo{author}{\bibfnamefont{D.~J.} \bibnamefont{Scalapino}},
  \bibinfo{journal}{Ann.~Phys.~(N.~Y.)} \textbf{\bibinfo{volume}{193}},
  \bibinfo{pages}{206} (\bibinfo{year}{1989}).

\bibitem[{\citenamefont{Monthoux and Scalapino}(1994)}]{monthoux94B}
\bibinfo{author}{\bibfnamefont{P.}~\bibnamefont{Monthoux}} \bibnamefont{and}
  \bibinfo{author}{\bibfnamefont{D.~J.} \bibnamefont{Scalapino}},
  \bibinfo{journal}{Phys. Rev. Lett.} \textbf{\bibinfo{volume}{72}},
  \bibinfo{pages}{1874} (\bibinfo{year}{1994}).

\bibitem[{\citenamefont{Pao and Bickers}(1994)}]{pao94}
\bibinfo{author}{\bibfnamefont{C.-H.} \bibnamefont{Pao}} \bibnamefont{and}
  \bibinfo{author}{\bibfnamefont{N.~E.} \bibnamefont{Bickers}},
  \bibinfo{journal}{Phys. Rev. Lett.} \textbf{\bibinfo{volume}{1870}},
  \bibinfo{pages}{72} (\bibinfo{year}{1994}).

\bibitem[{\citenamefont{Dahm and Tewordt}(1995)}]{dahm95}
\bibinfo{author}{\bibfnamefont{T.}~\bibnamefont{Dahm}} \bibnamefont{and}
  \bibinfo{author}{\bibfnamefont{L.}~\bibnamefont{Tewordt}},
  \bibinfo{journal}{Phys. Rev. Lett.} \textbf{\bibinfo{volume}{74}},
  \bibinfo{pages}{793} (\bibinfo{year}{1995}).

\bibitem[{\citenamefont{Scalapino}(1995)}]{scalapino95}
\bibinfo{author}{\bibfnamefont{D.~J.} \bibnamefont{Scalapino}},
  \bibinfo{journal}{Phys. Rep.} \textbf{\bibinfo{volume}{250}},
  \bibinfo{pages}{329} (\bibinfo{year}{1995}).

\bibitem[{\citenamefont{Lichtenstein et~al.}(1996)\citenamefont{Lichtenstein,
  Gunnarson, Andersen, and Martin}}]{lichtenstein96}
\bibinfo{author}{\bibfnamefont{A.~I.} \bibnamefont{Lichtenstein}},
  \bibinfo{author}{\bibfnamefont{O.}~\bibnamefont{Gunnarson}},
  \bibinfo{author}{\bibfnamefont{O.~K.} \bibnamefont{Andersen}},
  \bibnamefont{and} \bibinfo{author}{\bibfnamefont{R.~M.}
  \bibnamefont{Martin}}, \bibinfo{journal}{Phys. Rev. B}
  \textbf{\bibinfo{volume}{54}}, \bibinfo{pages}{12505} (\bibinfo{year}{1996}).

\bibitem[{\citenamefont{Manske et~al.}(2001)\citenamefont{Manske, Eremin, and
  Bennemann}}]{manske01}
\bibinfo{author}{\bibfnamefont{D.}~\bibnamefont{Manske}},
  \bibinfo{author}{\bibfnamefont{I.}~\bibnamefont{Eremin}}, \bibnamefont{and}
  \bibinfo{author}{\bibfnamefont{K.~H.} \bibnamefont{Bennemann}},
  \bibinfo{journal}{Phys. Rev. Lett.} \textbf{\bibinfo{volume}{87}},
  \bibinfo{pages}{177005} (\bibinfo{year}{2001}).

\bibitem[{\citenamefont{McMahan et~al.}(1988)\citenamefont{McMahan, Martin, and
  Satpathy}}]{mcmahan88}
\bibinfo{author}{\bibfnamefont{A.~K.} \bibnamefont{McMahan}},
  \bibinfo{author}{\bibfnamefont{R.~M.} \bibnamefont{Martin}},
  \bibnamefont{and} \bibinfo{author}{\bibfnamefont{S.}~\bibnamefont{Satpathy}},
  \bibinfo{journal}{Phys. Rev. B} \textbf{\bibinfo{volume}{38}},
  \bibinfo{pages}{6650} (\bibinfo{year}{1988}).

\bibitem[{\citenamefont{Hybertsen et~al.}(1989)\citenamefont{Hybertsen,
  Schl{\"u}ter, and Christensen}}]{hybertsen89}
\bibinfo{author}{\bibfnamefont{M.~S.} \bibnamefont{Hybertsen}},
  \bibinfo{author}{\bibfnamefont{M.}~\bibnamefont{Schl{\"u}ter}},
  \bibnamefont{and} \bibinfo{author}{\bibfnamefont{N.~E.}
  \bibnamefont{Christensen}}, \bibinfo{journal}{Phys. Rev. B}
  \textbf{\bibinfo{volume}{39}}, \bibinfo{pages}{9028} (\bibinfo{year}{1989}).

\bibitem[{\citenamefont{Monthoux et~al.}(1992)\citenamefont{Monthoux, Balatsky,
  and Pines}}]{monthoux92}
\bibinfo{author}{\bibfnamefont{P.}~\bibnamefont{Monthoux}},
  \bibinfo{author}{\bibfnamefont{A.~V.} \bibnamefont{Balatsky}},
  \bibnamefont{and} \bibinfo{author}{\bibfnamefont{D.}~\bibnamefont{Pines}},
  \bibinfo{journal}{Phys. Rev. B} \textbf{\bibinfo{volume}{46}},
  \bibinfo{pages}{14803} (\bibinfo{year}{1992}).

\bibitem[{\citenamefont{Monthoux and Pines}(1993)}]{monthoux93}
\bibinfo{author}{\bibfnamefont{P.}~\bibnamefont{Monthoux}} \bibnamefont{and}
  \bibinfo{author}{\bibfnamefont{D.}~\bibnamefont{Pines}},
  \bibinfo{journal}{Phys. Rev. B} \textbf{\bibinfo{volume}{47}},
  \bibinfo{pages}{6069} (\bibinfo{year}{1993}).

\bibitem[{\citenamefont{Monthoux and Pines}(1994)}]{monthoux94A}
\bibinfo{author}{\bibfnamefont{P.}~\bibnamefont{Monthoux}} \bibnamefont{and}
  \bibinfo{author}{\bibfnamefont{D.}~\bibnamefont{Pines}},
  \bibinfo{journal}{Phys. Rev. B} \textbf{\bibinfo{volume}{49}},
  \bibinfo{pages}{4261} (\bibinfo{year}{1994}).

\bibitem[{\citenamefont{Millis et~al.}(1990)\citenamefont{Millis, Monien, and
  Pines}}]{millis90}
\bibinfo{author}{\bibfnamefont{A.~J.} \bibnamefont{Millis}},
  \bibinfo{author}{\bibfnamefont{H.}~\bibnamefont{Monien}}, \bibnamefont{and}
  \bibinfo{author}{\bibfnamefont{D.}~\bibnamefont{Pines}},
  \bibinfo{journal}{Phys. Rev. B} \textbf{\bibinfo{volume}{42}},
  \bibinfo{pages}{167} (\bibinfo{year}{1990}).

\bibitem[{\citenamefont{Fong et~al.}(1996)\citenamefont{Fong, Keimer, Reznik,
  Milius, and Aksay}}]{fong96}
\bibinfo{author}{\bibfnamefont{H.~F.} \bibnamefont{Fong}},
  \bibinfo{author}{\bibfnamefont{B.}~\bibnamefont{Keimer}},
  \bibinfo{author}{\bibfnamefont{D.}~\bibnamefont{Reznik}},
  \bibinfo{author}{\bibfnamefont{D.~L.} \bibnamefont{Milius}},
  \bibnamefont{and} \bibinfo{author}{\bibfnamefont{I.~A.} \bibnamefont{Aksay}},
  \bibinfo{journal}{Phys. Rev. B} \textbf{\bibinfo{volume}{54}},
  \bibinfo{pages}{6708} (\bibinfo{year}{1996}).

\bibitem[{\citenamefont{Fong et~al.}(1999)\citenamefont{Fong, Bourges, Sidis,
  Regnault, Ivanov, Gu, Koshizuka, and Keimer}}]{fong99}
\bibinfo{author}{\bibfnamefont{H.~F.} \bibnamefont{Fong}},
  \bibinfo{author}{\bibfnamefont{P.}~\bibnamefont{Bourges}},
  \bibinfo{author}{\bibfnamefont{Y.}~\bibnamefont{Sidis}},
  \bibinfo{author}{\bibfnamefont{L.~P.} \bibnamefont{Regnault}},
  \bibinfo{author}{\bibfnamefont{A.}~\bibnamefont{Ivanov}},
  \bibinfo{author}{\bibfnamefont{G.~D.} \bibnamefont{Gu}},
  \bibinfo{author}{\bibfnamefont{N.}~\bibnamefont{Koshizuka}},
  \bibnamefont{and} \bibinfo{author}{\bibfnamefont{B.}~\bibnamefont{Keimer}},
  \bibinfo{journal}{Nature} \textbf{\bibinfo{volume}{398}},
  \bibinfo{pages}{588} (\bibinfo{year}{1999}).

\bibitem[{\citenamefont{Fong et~al.}(2000)\citenamefont{Fong, Bourges, Sidis,
  Regnault, Bossy, Ivanov, Milius, Aksay, and Keimer}}]{fong00}
\bibinfo{author}{\bibfnamefont{H.~F.} \bibnamefont{Fong}},
  \bibinfo{author}{\bibfnamefont{P.}~\bibnamefont{Bourges}},
  \bibinfo{author}{\bibfnamefont{Y.}~\bibnamefont{Sidis}},
  \bibinfo{author}{\bibfnamefont{L.~P.} \bibnamefont{Regnault}},
  \bibinfo{author}{\bibfnamefont{J.}~\bibnamefont{Bossy}},
  \bibinfo{author}{\bibfnamefont{A.}~\bibnamefont{Ivanov}},
  \bibinfo{author}{\bibfnamefont{D.~L.} \bibnamefont{Milius}},
  \bibinfo{author}{\bibfnamefont{I.~A.} \bibnamefont{Aksay}}, \bibnamefont{and}
  \bibinfo{author}{\bibfnamefont{B.}~\bibnamefont{Keimer}},
  \bibinfo{journal}{Phys. Rev. B} \textbf{\bibinfo{volume}{61}},
  \bibinfo{pages}{14773} (\bibinfo{year}{2000}).

\bibitem[{\citenamefont{He et~al.}(2002)\citenamefont{He, Bourges, Sidis,
  Ulrich, Regnault, Pailh{\`e}s, Berzigiarova, Kolesnikov, and Keimer}}]{he02}
\bibinfo{author}{\bibfnamefont{H.}~\bibnamefont{He}},
  \bibinfo{author}{\bibfnamefont{P.}~\bibnamefont{Bourges}},
  \bibinfo{author}{\bibfnamefont{Y.}~\bibnamefont{Sidis}},
  \bibinfo{author}{\bibfnamefont{C.}~\bibnamefont{Ulrich}},
  \bibinfo{author}{\bibfnamefont{L.~P.} \bibnamefont{Regnault}},
  \bibinfo{author}{\bibfnamefont{S.}~\bibnamefont{Pailh{\`e}s}},
  \bibinfo{author}{\bibfnamefont{N.~S.} \bibnamefont{Berzigiarova}},
  \bibinfo{author}{\bibfnamefont{N.~N.} \bibnamefont{Kolesnikov}},
  \bibnamefont{and} \bibinfo{author}{\bibfnamefont{B.}~\bibnamefont{Keimer}},
  \bibinfo{journal}{Science} \textbf{\bibinfo{volume}{295}},
  \bibinfo{pages}{1045} (\bibinfo{year}{2002}).

\bibitem[{\citenamefont{Eschrig and Norman}(2003)}]{eschrig03}
\bibinfo{author}{\bibfnamefont{M.}~\bibnamefont{Eschrig}} \bibnamefont{and}
  \bibinfo{author}{\bibfnamefont{M.}~\bibnamefont{Norman}},
  \bibinfo{journal}{Phys. Rev. B} \textbf{\bibinfo{volume}{67}},
  \bibinfo{pages}{144503} (\bibinfo{year}{2003}).

\bibitem[{\citenamefont{Damascelli et~al.}(2003)\citenamefont{Damascelli,
  Hussain, and Shen}}]{damascelli03}
\bibinfo{author}{\bibfnamefont{A.}~\bibnamefont{Damascelli}},
  \bibinfo{author}{\bibfnamefont{Z.}~\bibnamefont{Hussain}}, \bibnamefont{and}
  \bibinfo{author}{\bibfnamefont{Z.-X.} \bibnamefont{Shen}},
  \bibinfo{journal}{Rev. Mod. Phys.} \textbf{\bibinfo{volume}{75}},
  \bibinfo{pages}{473} (\bibinfo{year}{2003}).

\bibitem[{\citenamefont{Schachinger et~al.}(1997)\citenamefont{Schachinger,
  Carbotte, and Marsiglio}}]{schachinger97}
\bibinfo{author}{\bibfnamefont{E.}~\bibnamefont{Schachinger}},
  \bibinfo{author}{\bibfnamefont{J.~P.} \bibnamefont{Carbotte}},
  \bibnamefont{and}
  \bibinfo{author}{\bibfnamefont{F.}~\bibnamefont{Marsiglio}},
  \bibinfo{journal}{Phys. Rev. B} \textbf{\bibinfo{volume}{56}},
  \bibinfo{pages}{2738} (\bibinfo{year}{1997}).

\bibitem[{\citenamefont{Quinlan et~al.}(1996)\citenamefont{Quinlan, Hirschfeld,
  and Scalapino}}]{quinlan96}
\bibinfo{author}{\bibfnamefont{S.~M.} \bibnamefont{Quinlan}},
  \bibinfo{author}{\bibfnamefont{P.~J.} \bibnamefont{Hirschfeld}},
  \bibnamefont{and} \bibinfo{author}{\bibfnamefont{D.~J.}
  \bibnamefont{Scalapino}}, \bibinfo{journal}{Phys. Rev. B}
  \textbf{\bibinfo{volume}{53}}, \bibinfo{pages}{8575} (\bibinfo{year}{1996}).

\bibitem[{\citenamefont{Puchkov et~al.}(1996)\citenamefont{Puchkov, Basov, and
  Timusk}}]{puchkov96}
\bibinfo{author}{\bibfnamefont{A.~V.} \bibnamefont{Puchkov}},
  \bibinfo{author}{\bibfnamefont{D.~N.} \bibnamefont{Basov}}, \bibnamefont{and}
  \bibinfo{author}{\bibfnamefont{T.}~\bibnamefont{Timusk}},
  \bibinfo{journal}{J. Phys. Condens. Matter} \textbf{\bibinfo{volume}{8}},
  \bibinfo{pages}{10049} (\bibinfo{year}{1996}).

\bibitem[{\citenamefont{Bernhard et~al.}(2002)\citenamefont{Bernhard, Holden,
  Huml\'{i}\v{c}ek, Munzar, Golnik, Kl{\"a}ser, Wolf, Carr, Holmes, Keimer
  et~al.}}]{bernhard02}
\bibinfo{author}{\bibfnamefont{C.}~\bibnamefont{Bernhard}},
  \bibinfo{author}{\bibfnamefont{T.}~\bibnamefont{Holden}},
  \bibinfo{author}{\bibfnamefont{J.}~\bibnamefont{Huml\'{i}\v{c}ek}},
  \bibinfo{author}{\bibfnamefont{D.}~\bibnamefont{Munzar}},
  \bibinfo{author}{\bibfnamefont{A.}~\bibnamefont{Golnik}},
  \bibinfo{author}{\bibfnamefont{M.}~\bibnamefont{Kl{\"a}ser}},
  \bibinfo{author}{\bibfnamefont{T.}~\bibnamefont{Wolf}},
  \bibinfo{author}{\bibfnamefont{L.}~\bibnamefont{Carr}},
  \bibinfo{author}{\bibfnamefont{C.}~\bibnamefont{Holmes}},
  \bibinfo{author}{\bibfnamefont{B.}~\bibnamefont{Keimer}},
  \bibnamefont{et~al.}, \bibinfo{journal}{Solid State Commun.}
  \textbf{\bibinfo{volume}{121}}, \bibinfo{pages}{93} (\bibinfo{year}{2002}).

\bibitem[{\citenamefont{Tu et~al.}(2002)\citenamefont{Tu, Homes, Gu, Basov, and
  Strongin}}]{tu02}
\bibinfo{author}{\bibfnamefont{J.~J.} \bibnamefont{Tu}},
  \bibinfo{author}{\bibfnamefont{C.~C.} \bibnamefont{Homes}},
  \bibinfo{author}{\bibfnamefont{G.~D.} \bibnamefont{Gu}},
  \bibinfo{author}{\bibfnamefont{D.~N.} \bibnamefont{Basov}}, \bibnamefont{and}
  \bibinfo{author}{\bibfnamefont{M.}~\bibnamefont{Strongin}},
  \bibinfo{journal}{Phys. Rev. B} \textbf{\bibinfo{volume}{66}},
  \bibinfo{pages}{144514} (\bibinfo{year}{2002}).

\bibitem[{\citenamefont{van~der Marel et~al.}(2003)\citenamefont{van~der Marel,
  Molegraaf, Zaanen, Nussinov, Carbone, Damascelli, Eisaki, Greven, Kes, and
  Li}}]{vdmarel03}
\bibinfo{author}{\bibfnamefont{D.}~\bibnamefont{van~der Marel}},
  \bibinfo{author}{\bibfnamefont{H.~J.~A.} \bibnamefont{Molegraaf}},
  \bibinfo{author}{\bibfnamefont{J.}~\bibnamefont{Zaanen}},
  \bibinfo{author}{\bibfnamefont{Z.}~\bibnamefont{Nussinov}},
  \bibinfo{author}{\bibfnamefont{F.}~\bibnamefont{Carbone}},
  \bibinfo{author}{\bibfnamefont{A.}~\bibnamefont{Damascelli}},
  \bibinfo{author}{\bibfnamefont{H.}~\bibnamefont{Eisaki}},
  \bibinfo{author}{\bibfnamefont{M.}~\bibnamefont{Greven}},
  \bibinfo{author}{\bibfnamefont{P.~H.} \bibnamefont{Kes}}, \bibnamefont{and}
  \bibinfo{author}{\bibfnamefont{M.}~\bibnamefont{Li}},
  \bibinfo{journal}{Nature} \textbf{\bibinfo{volume}{425}},
  \bibinfo{pages}{271} (\bibinfo{year}{2003}).

\bibitem[{\citenamefont{Homes et~al.}(2004)\citenamefont{Homes, Dordevic, Bonn,
  Liang, and Hardy}}]{homes04}
\bibinfo{author}{\bibfnamefont{C.~C.} \bibnamefont{Homes}},
  \bibinfo{author}{\bibfnamefont{S.~V.} \bibnamefont{Dordevic}},
  \bibinfo{author}{\bibfnamefont{D.~A.} \bibnamefont{Bonn}},
  \bibinfo{author}{\bibfnamefont{R.}~\bibnamefont{Liang}}, \bibnamefont{and}
  \bibinfo{author}{\bibfnamefont{W.~N.} \bibnamefont{Hardy}},
  \bibinfo{journal}{Phys. Rev. B} \textbf{\bibinfo{volume}{69}},
  \bibinfo{pages}{024514} (\bibinfo{year}{2004}).

\bibitem[{\citenamefont{Boris et~al.}(2004)\citenamefont{Boris, Kovaleva,
  Dolgov, Lin, Keimer, and Bernhard}}]{boris04}
\bibinfo{author}{\bibfnamefont{A.~V.} \bibnamefont{Boris}},
  \bibinfo{author}{\bibfnamefont{N.~N.} \bibnamefont{Kovaleva}},
  \bibinfo{author}{\bibfnamefont{O.~V.} \bibnamefont{Dolgov}},
  \bibinfo{author}{\bibfnamefont{C.~T.} \bibnamefont{Lin}},
  \bibinfo{author}{\bibfnamefont{B.}~\bibnamefont{Keimer}}, \bibnamefont{and}
  \bibinfo{author}{\bibfnamefont{C.}~\bibnamefont{Bernhard}},
  \bibinfo{journal}{Science} \textbf{\bibinfo{volume}{121}},
  \bibinfo{pages}{93} (\bibinfo{year}{2004}).

\bibitem[{\citenamefont{Munzar et~al.}(1999)\citenamefont{Munzar, Bernhard, and
  Cardona}}]{munzar99}
\bibinfo{author}{\bibfnamefont{D.}~\bibnamefont{Munzar}},
  \bibinfo{author}{\bibfnamefont{C.}~\bibnamefont{Bernhard}}, \bibnamefont{and}
  \bibinfo{author}{\bibfnamefont{M.}~\bibnamefont{Cardona}},
  \bibinfo{journal}{Physica C} \textbf{\bibinfo{volume}{312}},
  \bibinfo{pages}{121} (\bibinfo{year}{1999}).

\bibitem[{\citenamefont{Carbotte et~al.}(1999)\citenamefont{Carbotte,
  Schachinger, and Basov}}]{carbotte99}
\bibinfo{author}{\bibfnamefont{J.~P.} \bibnamefont{Carbotte}},
  \bibinfo{author}{\bibfnamefont{E.}~\bibnamefont{Schachinger}},
  \bibnamefont{and} \bibinfo{author}{\bibfnamefont{D.~N.} \bibnamefont{Basov}},
  \bibinfo{journal}{Nature} \textbf{\bibinfo{volume}{401}},
  \bibinfo{pages}{354} (\bibinfo{year}{1999}).

\bibitem[{Sch({\natexlab{a}})}]{Schulga}
\bibinfo{note}{The relation can be easily obtained using the formalism
  summarized in the lecture notes of S.~V.~Schulga: S.~V.~Schulga, in {\it
  Material Science, Fundamental Properties and Future Electronic Applications
  of High-Tc Superconductors}, edited by S.~L.~Drechsler and T.~Mischonov
  (Kluwer Academic, Dordrecht, 2001), pp.~323-360; also published as
  cond-mat/0101243.}

\bibitem[{\citenamefont{Abanov et~al.}(2001{\natexlab{a}})\citenamefont{Abanov,
  Chubukov, and Schmalian}}]{abanov01A}
\bibinfo{author}{\bibfnamefont{A.}~\bibnamefont{Abanov}},
  \bibinfo{author}{\bibfnamefont{A.~V.} \bibnamefont{Chubukov}},
  \bibnamefont{and}
  \bibinfo{author}{\bibfnamefont{J.}~\bibnamefont{Schmalian}},
  \bibinfo{journal}{Phys. Rev. B} \textbf{\bibinfo{volume}{63}},
  \bibinfo{pages}{180510} (\bibinfo{year}{2001}{\natexlab{a}}).

\bibitem[{\citenamefont{Abanov et~al.}(2001{\natexlab{b}})\citenamefont{Abanov,
  Chubukov, and Schmalian}}]{abanov01B}
\bibinfo{author}{\bibfnamefont{A.}~\bibnamefont{Abanov}},
  \bibinfo{author}{\bibfnamefont{A.~V.} \bibnamefont{Chubukov}},
  \bibnamefont{and}
  \bibinfo{author}{\bibfnamefont{J.}~\bibnamefont{Schmalian}},
  \bibinfo{journal}{J.~Electron Spectrosc.} \textbf{\bibinfo{volume}{117}},
  \bibinfo{pages}{129} (\bibinfo{year}{2001}{\natexlab{b}}).

\bibitem[{\citenamefont{Schriefer}(1988)}]{schrieffer}
\bibinfo{author}{\bibfnamefont{J.~R.} \bibnamefont{Schriefer}},
  \emph{\bibinfo{title}{Theory of Superconductivity}}
  (\bibinfo{publisher}{Addison-Wesley, Reading}, \bibinfo{year}{1988}).

\bibitem[{\citenamefont{Sandvik et~al.}(2004)\citenamefont{Sandvik, Scalapino,
  and Bickers}}]{sandvik04}
\bibinfo{author}{\bibfnamefont{A.~W.} \bibnamefont{Sandvik}},
  \bibinfo{author}{\bibfnamefont{D.~J.} \bibnamefont{Scalapino}},
  \bibnamefont{and} \bibinfo{author}{\bibfnamefont{N.~E.}
  \bibnamefont{Bickers}}, \bibinfo{journal}{Phys. Rev. B}
  \textbf{\bibinfo{volume}{69}}, \bibinfo{pages}{094523}
  (\bibinfo{year}{2004}).

\bibitem[{\citenamefont{Schachinger et~al.}(2003)\citenamefont{Schachinger, Tu,
  and Carbotte}}]{schachinger03}
\bibinfo{author}{\bibfnamefont{E.}~\bibnamefont{Schachinger}},
  \bibinfo{author}{\bibfnamefont{J.~J.} \bibnamefont{Tu}}, \bibnamefont{and}
  \bibinfo{author}{\bibfnamefont{J.~P.} \bibnamefont{Carbotte}},
  \bibinfo{journal}{Phys. Rev. B} \textbf{\bibinfo{volume}{67}},
  \bibinfo{pages}{214508} (\bibinfo{year}{2003}).

\bibitem[{\citenamefont{Molegraaf et~al.}(2002)\citenamefont{Molegraaf,
  Presura, der Marel, Kes, and Li}}]{molegraaf02}
\bibinfo{author}{\bibfnamefont{H.~J.~A.} \bibnamefont{Molegraaf}},
  \bibinfo{author}{\bibfnamefont{C.}~\bibnamefont{Presura}},
  \bibinfo{author}{\bibfnamefont{D.~V.} \bibnamefont{der Marel}},
  \bibinfo{author}{\bibfnamefont{P.~H.} \bibnamefont{Kes}}, \bibnamefont{and}
  \bibinfo{author}{\bibfnamefont{M.}~\bibnamefont{Li}},
  \bibinfo{journal}{Science} \textbf{\bibinfo{volume}{295}},
  \bibinfo{pages}{2239} (\bibinfo{year}{2002}).

\bibitem[{Mah({\natexlab{a}})}]{Mahan1}
\bibinfo{note}{G.~D.~Mahan, {\it Many Particle Physics} (Kluver Academic/Plenum
  Publishers, New York, 2000), Chapt.~3.8.}

\bibitem[{\citenamefont{Scalapino et~al.}(1992)\citenamefont{Scalapino, White,
  and Zhang}}]{scalapino92}
\bibinfo{author}{\bibfnamefont{D.~J.} \bibnamefont{Scalapino}},
  \bibinfo{author}{\bibfnamefont{S.~R.} \bibnamefont{White}}, \bibnamefont{and}
  \bibinfo{author}{\bibfnamefont{S.~C.} \bibnamefont{Zhang}},
  \bibinfo{journal}{Phys. Rev. Lett.} \textbf{\bibinfo{volume}{68}},
  \bibinfo{pages}{2830} (\bibinfo{year}{1992}).

\bibitem[{Sch({\natexlab{b}})}]{Schrieffer2}
\bibinfo{note}{For a derivation of the formula see Chapt.~8 of of
  Ref.~\onlinecite{schrieffer}.}

\bibitem[{Mah({\natexlab{b}})}]{Mahan2}
\bibinfo{note}{Eq.~\ref{eq:cond} has been derived along the lines of
  Chapt.~8.1.2 of Ref.~\onlinecite{Mahan1}.}

\bibitem[{\citenamefont{Andersen et~al.}(1995)\citenamefont{Andersen,
  Liechtenstein, Jepsen, and Paulsen}}]{andersen95}
\bibinfo{author}{\bibfnamefont{O.~K.} \bibnamefont{Andersen}},
  \bibinfo{author}{\bibfnamefont{A.~I.} \bibnamefont{Liechtenstein}},
  \bibinfo{author}{\bibfnamefont{O.}~\bibnamefont{Jepsen}}, \bibnamefont{and}
  \bibinfo{author}{\bibfnamefont{F.}~\bibnamefont{Paulsen}},
  \bibinfo{journal}{J. Phys. Chem. Solids} \textbf{\bibinfo{volume}{56}},
  \bibinfo{pages}{1573} (\bibinfo{year}{1995}).

\bibitem[{\citenamefont{Feng et~al.}(2001)\citenamefont{Feng, Armitage, Lu,
  Damascelli, Hu, Bogdanov, Lanzara, Ronning, Shen, Eisaki et~al.}}]{feng01}
\bibinfo{author}{\bibfnamefont{D.~L.} \bibnamefont{Feng}},
  \bibinfo{author}{\bibfnamefont{N.~D.} \bibnamefont{Armitage}},
  \bibinfo{author}{\bibfnamefont{D.~H.} \bibnamefont{Lu}},
  \bibinfo{author}{\bibfnamefont{A.}~\bibnamefont{Damascelli}},
  \bibinfo{author}{\bibfnamefont{J.~P.} \bibnamefont{Hu}},
  \bibinfo{author}{\bibfnamefont{P.}~\bibnamefont{Bogdanov}},
  \bibinfo{author}{\bibfnamefont{A.}~\bibnamefont{Lanzara}},
  \bibinfo{author}{\bibfnamefont{F.}~\bibnamefont{Ronning}},
  \bibinfo{author}{\bibfnamefont{K.~M.} \bibnamefont{Shen}},
  \bibinfo{author}{\bibfnamefont{H.}~\bibnamefont{Eisaki}},
  \bibnamefont{et~al.}, \bibinfo{journal}{Phys. Rev. Lett.}
  \textbf{\bibinfo{volume}{86}}, \bibinfo{pages}{5550} (\bibinfo{year}{2001}).

\bibitem[{\citenamefont{Chuang et~al.}(2001)\citenamefont{Chuang, Gromko,
  Fedorov, Aiura, Oka, Ando, Eisaki, Uchida, and Dessau}}]{chuang01}
\bibinfo{author}{\bibfnamefont{Y.~D.} \bibnamefont{Chuang}},
  \bibinfo{author}{\bibfnamefont{A.~D.} \bibnamefont{Gromko}},
  \bibinfo{author}{\bibfnamefont{A.}~\bibnamefont{Fedorov}},
  \bibinfo{author}{\bibfnamefont{Y.}~\bibnamefont{Aiura}},
  \bibinfo{author}{\bibfnamefont{K.}~\bibnamefont{Oka}},
  \bibinfo{author}{\bibfnamefont{Y.}~\bibnamefont{Ando}},
  \bibinfo{author}{\bibfnamefont{H.}~\bibnamefont{Eisaki}},
  \bibinfo{author}{\bibfnamefont{S.~I.} \bibnamefont{Uchida}},
  \bibnamefont{and} \bibinfo{author}{\bibfnamefont{D.~S.}
  \bibnamefont{Dessau}}, \bibinfo{journal}{Phys. Rev. Lett.}
  \textbf{\bibinfo{volume}{87}}, \bibinfo{pages}{117002}
  (\bibinfo{year}{2001}).

\bibitem[{\citenamefont{Kordyuk et~al.}(2002)\citenamefont{Kordyuk, Borisenko,
  Kim, Nenkov, Knupfer, Fink, Golden, Berger, and Follath}}]{kordyuk02}
\bibinfo{author}{\bibfnamefont{A.~A.} \bibnamefont{Kordyuk}},
  \bibinfo{author}{\bibfnamefont{S.~V.} \bibnamefont{Borisenko}},
  \bibinfo{author}{\bibfnamefont{T.~K.} \bibnamefont{Kim}},
  \bibinfo{author}{\bibfnamefont{K.~A.} \bibnamefont{Nenkov}},
  \bibinfo{author}{\bibfnamefont{M.}~\bibnamefont{Knupfer}},
  \bibinfo{author}{\bibfnamefont{J.}~\bibnamefont{Fink}},
  \bibinfo{author}{\bibfnamefont{M.~S.} \bibnamefont{Golden}},
  \bibinfo{author}{\bibfnamefont{H.}~\bibnamefont{Berger}}, \bibnamefont{and}
  \bibinfo{author}{\bibfnamefont{R.}~\bibnamefont{Follath}},
  \bibinfo{journal}{Phys. Rev. Lett.} \textbf{\bibinfo{volume}{89}},
  \bibinfo{pages}{077003} (\bibinfo{year}{2002}).

\bibitem[{\citenamefont{Eschrig and Norman}(2002)}]{eschrig02}
\bibinfo{author}{\bibfnamefont{M.}~\bibnamefont{Eschrig}} \bibnamefont{and}
  \bibinfo{author}{\bibfnamefont{M.~R.} \bibnamefont{Norman}},
  \bibinfo{journal}{Phys. Rev. Lett.} \textbf{\bibinfo{volume}{89}},
  \bibinfo{pages}{277005} (\bibinfo{year}{2002}).

\bibitem[{\citenamefont{Borisenko et~al.}(2003)\citenamefont{Borisenko,
  Kordyuk, Kim, Koitzsch, Knupfer, Fink, Golden, Eschrig, Berger, and
  Follath}}]{borisenko03}
\bibinfo{author}{\bibfnamefont{S.~V.} \bibnamefont{Borisenko}},
  \bibinfo{author}{\bibfnamefont{A.~A.} \bibnamefont{Kordyuk}},
  \bibinfo{author}{\bibfnamefont{T.~K.} \bibnamefont{Kim}},
  \bibinfo{author}{\bibfnamefont{A.}~\bibnamefont{Koitzsch}},
  \bibinfo{author}{\bibfnamefont{M.}~\bibnamefont{Knupfer}},
  \bibinfo{author}{\bibfnamefont{J.}~\bibnamefont{Fink}},
  \bibinfo{author}{\bibfnamefont{M.~S.} \bibnamefont{Golden}},
  \bibinfo{author}{\bibfnamefont{M.}~\bibnamefont{Eschrig}},
  \bibinfo{author}{\bibfnamefont{H.}~\bibnamefont{Berger}}, \bibnamefont{and}
  \bibinfo{author}{\bibfnamefont{R.}~\bibnamefont{Follath}},
  \bibinfo{journal}{Phys. Rev. Lett.} \textbf{\bibinfo{volume}{90}},
  \bibinfo{pages}{207001} (\bibinfo{year}{2003}).

\bibitem[{li0()}]{li05}
\bibinfo{note}{Jian-Xin Li, T. Zhou, and Z. D. Wang, cond-mat/0501356
  (unpublished)}.

\bibitem[{\citenamefont{Pailhes et~al.}(2003)\citenamefont{Pailhes, Sidis,
  Bourges, Ulrich, Hinkov, Regnault, Ivanov, Liang, Lin, Bernhard
  et~al.}}]{pailhes03}
\bibinfo{author}{\bibfnamefont{S.}~\bibnamefont{Pailhes}},
  \bibinfo{author}{\bibfnamefont{Y.}~\bibnamefont{Sidis}},
  \bibinfo{author}{\bibfnamefont{P.}~\bibnamefont{Bourges}},
  \bibinfo{author}{\bibfnamefont{C.}~\bibnamefont{Ulrich}},
  \bibinfo{author}{\bibfnamefont{V.}~\bibnamefont{Hinkov}},
  \bibinfo{author}{\bibfnamefont{L.~P.} \bibnamefont{Regnault}},
  \bibinfo{author}{\bibfnamefont{A.}~\bibnamefont{Ivanov}},
  \bibinfo{author}{\bibfnamefont{B.}~\bibnamefont{Liang}},
  \bibinfo{author}{\bibfnamefont{C.~T.} \bibnamefont{Lin}},
  \bibinfo{author}{\bibfnamefont{C.}~\bibnamefont{Bernhard}},
  \bibnamefont{et~al.}, \bibinfo{journal}{Phys. Rev. Lett.}
  \textbf{\bibinfo{volume}{91}}, \bibinfo{pages}{237002}
  (\bibinfo{year}{2003}).

\bibitem[{\citenamefont{Kee et~al.}(2002)\citenamefont{Kee, Kivelson, and
  Aeppli}}]{kee02}
\bibinfo{author}{\bibfnamefont{H.~Y.} \bibnamefont{Kee}},
  \bibinfo{author}{\bibfnamefont{S.~A.} \bibnamefont{Kivelson}},
  \bibnamefont{and} \bibinfo{author}{\bibfnamefont{G.}~\bibnamefont{Aeppli}},
  \bibinfo{journal}{Phys. Rev. Lett.} \textbf{\bibinfo{volume}{88}},
  \bibinfo{pages}{257002} (\bibinfo{year}{2002}).

\bibitem[{\citenamefont{Abanov et~al.}(2002)\citenamefont{Abanov, Chubukov,
  Eschrig, Norman, and Schmalian}}]{abanov02}
\bibinfo{author}{\bibfnamefont{A.}~\bibnamefont{Abanov}},
  \bibinfo{author}{\bibfnamefont{A.~V.} \bibnamefont{Chubukov}},
  \bibinfo{author}{\bibfnamefont{M.}~\bibnamefont{Eschrig}},
  \bibinfo{author}{\bibfnamefont{M.~R.} \bibnamefont{Norman}},
  \bibnamefont{and}
  \bibinfo{author}{\bibfnamefont{J.}~\bibnamefont{Schmalian}},
  \bibinfo{journal}{Phys. Rev. Lett.} \textbf{\bibinfo{volume}{89}},
  \bibinfo{pages}{177002} (\bibinfo{year}{2002}).

\bibitem[{esc()}]{eschrigg}
\bibinfo{note}{Note that the authors of Ref.~\onlinecite{eschrig03} use the
  spin susceptibility normalized to the experimental value of the spectral
  weight of the mode. Furthermore, their formula for the selfenergy does not
  contain the factor of $3/4$ of Eq.~\ref{eq:Selfen3}. The effective value of
  $g$ corresponding to their spin susceptibility normalized according to
  Eq.~\ref{eq:normalizationchi} and our selfenergy formula is about
  $0.25{\rm\,eV}$.}

\bibitem[{\citenamefont{Hwang et~al.}(2004)\citenamefont{Hwang, Timusk, and
  Gu}}]{hwang04}
\bibinfo{author}{\bibfnamefont{J.}~\bibnamefont{Hwang}},
  \bibinfo{author}{\bibfnamefont{T.}~\bibnamefont{Timusk}}, \bibnamefont{and}
  \bibinfo{author}{\bibfnamefont{G.~D.} \bibnamefont{Gu}},
  \bibinfo{journal}{Nature} \textbf{\bibinfo{volume}{295}},
  \bibinfo{pages}{1045} (\bibinfo{year}{2004}).

\bibitem[{\citenamefont{Schachinger and Carbotte}(2000)}]{schachinger00}
\bibinfo{author}{\bibfnamefont{E.}~\bibnamefont{Schachinger}} \bibnamefont{and}
  \bibinfo{author}{\bibfnamefont{J.~P.} \bibnamefont{Carbotte}},
  \bibinfo{journal}{Phys. Rev. B} \textbf{\bibinfo{volume}{62}},
  \bibinfo{pages}{9054} (\bibinfo{year}{2000}).

\bibitem[{\citenamefont{Singley et~al.}(2001)\citenamefont{Singley, Basov,
  Kurahashi, Uefuji, and Yamada}}]{singley01}
\bibinfo{author}{\bibfnamefont{E.~J.} \bibnamefont{Singley}},
  \bibinfo{author}{\bibfnamefont{D.~N.} \bibnamefont{Basov}},
  \bibinfo{author}{\bibfnamefont{K.}~\bibnamefont{Kurahashi}},
  \bibinfo{author}{\bibfnamefont{T.}~\bibnamefont{Uefuji}}, \bibnamefont{and}
  \bibinfo{author}{\bibfnamefont{K.}~\bibnamefont{Yamada}},
  \bibinfo{journal}{Phys. Rev. B} \textbf{\bibinfo{volume}{64}},
  \bibinfo{pages}{224503} (\bibinfo{year}{2001}).

\bibitem[{\citenamefont{Wang et~al.}(2003)\citenamefont{Wang, Zheng, Luo, Chen,
  Yan, Fang, and Ma}}]{wang03}
\bibinfo{author}{\bibfnamefont{N.~L.} \bibnamefont{Wang}},
  \bibinfo{author}{\bibfnamefont{P.}~\bibnamefont{Zheng}},
  \bibinfo{author}{\bibfnamefont{J.~L.} \bibnamefont{Luo}},
  \bibinfo{author}{\bibfnamefont{Z.~J.} \bibnamefont{Chen}},
  \bibinfo{author}{\bibfnamefont{S.~L.} \bibnamefont{Yan}},
  \bibinfo{author}{\bibfnamefont{L.}~\bibnamefont{Fang}}, \bibnamefont{and}
  \bibinfo{author}{\bibfnamefont{Y.~C.} \bibnamefont{Ma}},
  \bibinfo{journal}{Phys. Rev. B} \textbf{\bibinfo{volume}{68}},
  \bibinfo{pages}{054516} (\bibinfo{year}{2003}).

\bibitem[{dor()}]{dordevic04}
\bibinfo{note}{S. V. Dordevic and C. C. Homes and J. J. Tu and T. Valla and M.
  Strongin and P. D. Johnson and G. D. Gu and D. N. Basov, cond-mat/0411043
  (unpublished).}

\bibitem[{eps()}]{epsinfty}
\bibinfo{note}{In order to obtain the scattering rate spectra from optical data
  one has to make an assumption regarding the interband contribution to the
  dielectric function. This brings about a certain degree of arbitrariness.}

\bibitem[{\citenamefont{Shen et~al.}(1995)\citenamefont{Shen, Spicer, King,
  Dessau, and Wells}}]{shen95}
\bibinfo{author}{\bibfnamefont{Z.~X.} \bibnamefont{Shen}},
  \bibinfo{author}{\bibfnamefont{W.~E.} \bibnamefont{Spicer}},
  \bibinfo{author}{\bibfnamefont{D.~M.} \bibnamefont{King}},
  \bibinfo{author}{\bibfnamefont{D.~S.} \bibnamefont{Dessau}},
  \bibnamefont{and} \bibinfo{author}{\bibfnamefont{B.~O.} \bibnamefont{Wells}},
  \bibinfo{journal}{Science} \textbf{\bibinfo{volume}{267}},
  \bibinfo{pages}{343} (\bibinfo{year}{1995}).

\bibitem[{\citenamefont{Millis and Drew}(2003)}]{drew04}
\bibinfo{author}{\bibfnamefont{A.~J.} \bibnamefont{Millis}} \bibnamefont{and}
  \bibinfo{author}{\bibfnamefont{H.~D.} \bibnamefont{Drew}},
  \bibinfo{journal}{Phys. Rev. B} \textbf{\bibinfo{volume}{67}},
  \bibinfo{pages}{214517} (\bibinfo{year}{2003}).

\bibitem[{mil()}]{millis04}
\bibinfo{note}{A.~J.~Millis, A.~Zimmers, R.~P.~S.~M.~Lobo, and N.~Bontemps,
  cond-mat/0411172 (unpublished).}

\bibitem[{\citenamefont{Norman et~al.}(2000)\citenamefont{Norman, Randeira,
  Janko, and Campuzano}}]{norman00}
\bibinfo{author}{\bibfnamefont{M.~R.} \bibnamefont{Norman}},
  \bibinfo{author}{\bibfnamefont{M.}~\bibnamefont{Randeira}},
  \bibinfo{author}{\bibfnamefont{B.}~\bibnamefont{Janko}}, \bibnamefont{and}
  \bibinfo{author}{\bibfnamefont{J.~C.} \bibnamefont{Campuzano}},
  \bibinfo{journal}{Phys. Rev. B} \textbf{\bibinfo{volume}{61}},
  \bibinfo{pages}{14742} (\bibinfo{year}{2000}).

\bibitem[{\citenamefont{Norman and P{\'e}pin}(2002)}]{norman02}
\bibinfo{author}{\bibfnamefont{M.~R.} \bibnamefont{Norman}} \bibnamefont{and}
  \bibinfo{author}{\bibfnamefont{C.}~\bibnamefont{P{\'e}pin}},
  \bibinfo{journal}{Phys. Rev. B} \textbf{\bibinfo{volume}{66}},
  \bibinfo{pages}{100506} (\bibinfo{year}{2002}).

\bibitem[{yan()}]{yanase04}
\bibinfo{note}{Y.~Yanase and M.~Ogata, cond-mat/0412508 (unpublished).}

\end{thebibliography}
\end{document}